\pdfoutput=1

\documentclass[conference]{IEEEtran}
\newcommand{\ignore}[1]{}
\usepackage[pass]{geometry}
\usepackage{fancyhdr}
\usepackage[normalem]{ulem}
\usepackage{soul}
\usepackage{longtable}
\usepackage[hyphens]{url}
\usepackage{hyperref}
\usepackage{amsmath}

\usepackage[boxed,ruled]{algorithm2e}
\usepackage[numbers,sort&compress]{natbib}
\usepackage{makecell}
\usepackage[para,online,flushleft]{threeparttable}
\usepackage{amssymb}
\usepackage{color}
\usepackage{caption}
\usepackage{graphicx}
\usepackage{placeins}
\usepackage{float}
\usepackage{subfigure}
\usepackage{multicol}
\usepackage{multirow}
\usepackage{fancyhdr}
\usepackage{xcolor}
\usepackage{amsmath}
\usepackage{mathtools}
\usepackage{xcolor}
\usepackage{soul}
\usepackage{comment}

\usepackage{color, soul}
\soulregister\cite7
\soulregister\ref7
\soulregister\underline7

\renewcommand\baselinestretch{0.92}

\soulregister\cite7
\soulregister\ref7
\soulregister\underline7

\usepackage{booktabs}
\newcommand{\tabincell}[2]{\begin{tabular}{@{}#1@{}}#2\end{tabular}}

\def\BibTeX{{\rm B\kern-.05em{\sc i\kern-.025em b}\kern-.08em
    T\kern-.1667em\lower.7ex\hbox{E}\kern-.125emX}}

\newcommand{\iscasubmissionnumber}{786}
\fancypagestyle{firstpage}{
  \fancyhf{}

  \fancyhead[C]{\normalsize{ISCA 2021 Submission
      \textbf{\#\iscasubmissionnumber} \\ Confidential Draft: DO NOT DISTRIBUTE}} 
  \fancyfoot[C]{\thepage}
}

\title{FORMS: Fine-grained Polarized ReRAM-based In-situ Computation for Mixed-signal DNN Accelerator} 

\author{
    Geng Yuan\text{$^\star$}\textsuperscript{\rm 1}\thanks{$^\star$These Authors contributed equally.},
    Payman Behnam\text{$^\star$}\textsuperscript{\rm 2},
    Zhengang Li\textsuperscript{\rm 1},
    Ali Shafiee\textsuperscript{\rm 3},
    Sheng Lin\textsuperscript{\rm 1},
    Xiaolong Ma\textsuperscript{\rm 1},
    Hang Liu\textsuperscript{\rm 4},\\
    Xuehai Qian\textsuperscript{\rm 5},
    Mahdi Nazm Bojnordi\textsuperscript{\rm 6},
    Yanzhi Wang\textsuperscript{\rm 1},
    Caiwen Ding\textsuperscript{\rm 7}\\
    \itshape\small
    \textsuperscript{1}Northeastern University, 
    \textsuperscript{2}Georgia Institute of Technology, 
    \textsuperscript{3}Samsung,
    \textsuperscript{4}Stevens Institute of Technology,\\
    \textsuperscript{5}University of Southern California,
    \itshape\small
    \textsuperscript{6}University of Utah,
    \textsuperscript{7}University of Connecticut\\
    
    {\tt\small \textsuperscript{\rm 1}\{yuan.geng, li.zhen, lin.sheng, ma.xiaol, yanz.wang\}@northeastern.edu,} \\
    {\tt\small \textsuperscript{\rm 2}payman.behnam@gatech.edu, }
    {\tt\small \textsuperscript{\rm 3}ali.shafiee@samsung.com, }
    {\tt\small \textsuperscript{\rm 4}hang.liu@stevens.edu,} \\
    {\tt\small \textsuperscript{\rm 5}xuehai.qian@usc.edu, }
    {\tt\small \textsuperscript{\rm 6}bojnordi@cs.utah.edu,}
    {\tt\small \textsuperscript{\rm 7}caiwen.ding@uconn.edu}
}
\IEEEoverridecommandlockouts

\begin{document}

\maketitle
% \thispagestyle{firstpage}
% \pagestyle{plain}

%%%%%% -- PAPER CONTENT STARTS-- %%%%%%%%

\begin{abstract}
Recent work demonstrated the promise of using 
resistive random access memory (ReRAM) as an 
emerging technology to perform inherently parallel
analog domain in-situ matrix-vector multiplication---the intensive and key computation in deep neural networks (DNNs).
One key problem is the weights that are signed values. 
However, in a ReRAM crossbar, weights are stored as conductance of the crossbar cells, and the in-situ computation assumes all cells on each crossbar column are of the same sign.
The current architectures either use two ReRAM crossbars
for positive and negative weights (PRIME), or add an offset
to weights so that all values become positive (ISAAC).
Neither solution is ideal: they either double the cost
of crossbars, or incur extra offset circuity. To better address this problem, we propose
{\em FORMS}, a fine-grained ReRAM-based DNN accelerator with algorithm/hardware co-design.
Instead of trying to represent the positive/negative weights,
our key design principle is to {\em enforce} exactly
what is assumed in the in-situ computation---ensuring that all weights in the same column of a crossbar 
have the {\em same sign}.
It naturally avoids the cost of an additional 
crossbar. 
Such polarized weights can be nicely generated using  
alternating direction method of multipliers (ADMM) regularized optimization during the DNN training, which can exactly 
enforce certain patterns in DNN weights.
To achieve high accuracy, we divide the crossbar
into logical sub-arrays and only enforce this property
within the fine-grained sub-array columns. 
Crucially, the small sub-arrays provides a {\em unique} opportunity for {\em input zero-skipping}, which can significantly avoid unnecessary computations and reduce computation time.
At the same time, it also makes the hardware much easier to implement and is less susceptible to non-idealities and noise than coarse-grained architectures.
Putting all together, 
with the same optimized DNN models, FORMS achieves
1.50$\times$ and 1.93$\times$ throughput improvement in terms of $\frac {GOPs}{s \times mm^{2}}$ and $\frac {GOPs}{W}$ compared to ISAAC, and
1.12$\times$ $\sim$ 2.4$\times$ speed up in terms of frame per second over optimized ISAAC with almost the same power/area cost.
Interestingly, FORMS optimization framework can even 
speed up the original ISAAC from 10.7$\times$ up to 377.9$\times$, reflecting the importance of software/hardware co-design optimizations.
\end{abstract}

\section{Introduction}
\label{intro}
Deep Neural Networks (DNNs) have become the fundamental element and core enabler of ubiquitous artificial intelligence, thanks to their high accuracy, excellent scalability, and self-adaptiveness~\cite{goodfellow2016deep}.
As the ever-growing DNN model size,
the high computation and memory storage of DNN models introduce substantial data movements, posing key challenges to the conventional Von Neumann architectures, where weight storage and computation units are separated. 
To reduce data movement, model compression techniques~\cite{han2015learning,wen2016learning,wu2016quantized} and hardware accelerators~\cite{chen2014diannao,tpu,ma2018area,ding2017circnn,wang2018towards,ding2018structured,yuan2019sot} have been intensively investigated. However, as Moore's law is reaching an end~\cite{waldrop2016chips}, the
potential of the acceleration architecture based on conventional technology
is still limited. 
We argue that drastic improvements can be only achieved by
1) the next-generation emerging device/circuit technology beyond 
CMOS; and 2) the vertical integration~\cite{cacm_patterson} and optimization of algorithm, architecture, and technology innovations to deliver better overall performance and energy efficiency 
for various applications. 

A promising emerging technology is the recently 
discovered resistive random access memory (ReRAM)~\cite{xia2017mnsim,thomas2013memristor} devices that are able to perform the inherently parallel
in-situ matrix-vector multiplication in the analog domain. 
This key feature has been applied to several significant problems, 
including solving systems of linear equations in $O(1)$ time complexity~\cite{chua1971memristor}, and more interestingly,
building DNN accelerators~\cite{PRIME,shafiee2016isaac, nag2018newton, song2017pipelayer, ankit2019puma, marinella2018multiscale, li2020timely, ghodrati2019mixed}.
Since the key computation in DNNs can be essentially expressed 
as matrix-vector multiplication, ReRAM crossbars can naturally
accelerate DNNs with much less data movement and low-cost computation. 
Due to their promising results, 
structured pruning~\cite{wen2016learning} and quantization~\cite{wu2016quantized} 
are also developed to facilitate a smaller number of weights and bits
to reduce the amount of computation and ReRAM hardware resources.

\begin{figure*} [t]
     \centering
%      \vspace{-1.2em}
     \includegraphics[width=2.0\columnwidth]{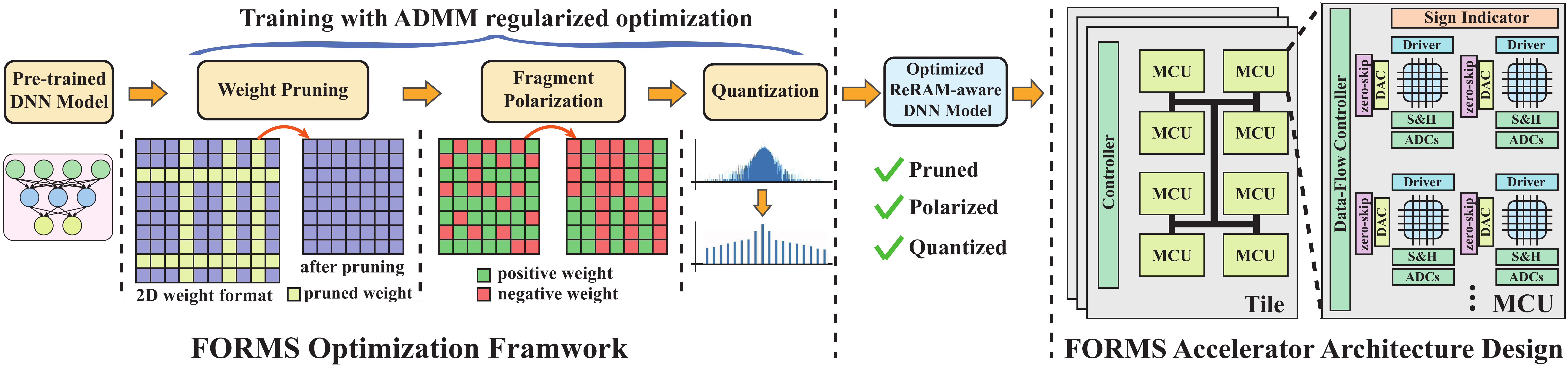} 
     \caption{Overall Flow of FORMS Algorithm/Hardware Co-designed.}
     \label{fig:overall_flow}
\end{figure*}

On the other side, a key complication of the ReRAM-based DNN accelerator
is that although the weights stored in ReRAM crossbar cells can be either
positive or negative, the in-situ computation assumes
all cells in each crossbar column hold values of the same sign, i.e.,
all positive or all negative. There are two approaches to tackle
the problem. 
The general way is to use two ReRAM crossbars to hold
the positive and negative magnitudes weights separately, doubling the ReRAM portion of hardware cost \cite{choi2018self,li2018efficient,liu2020neural, krestinskaya2020memristive, PRIME}.
In contrast, ISAAC~\cite{shafiee2016isaac} adds an offset to weights so that 
all values become positive. 
While keeping the number of crossbars the same, 
the latter approach 
introduces additional hardware costs for 
the peripheral circuits by adding extra offset circuits and may also decrease the network robustness to hardware failures~\cite{yuan2021failure}.
We argue that both solutions are not ideal, and we attempt to 
develop an alternative approach with better cost/performance trade-offs.

Different from the previous approaches, which use additional hardware
to ``fix'' the problem, our design principle is to {\em enforce}
exactly what is assumed in the in-situ computation---ensuring the 
{\em pattern} that all weights in the same column of a crossbar have the
{\em same sign}. 
This idea takes the opportunity of algorithm
and hardware co-design, and is motivated by the capability of the powerful alternating direction method of multipliers (ADMM) regularized optimization~\cite{boyd2011distributed}, which is exactly able to enforce patterns in DNN training while maintaining high accuracy.
Based on this idea, we can train a DNN model using ADMM with our novel constraints
such that the weights mapped to the same crossbar columns are all 
positive or negative. 
With the typical ReRAM crossbar size, e.g., $128 \times 128$,
we found that the ``same-sign'' property of the coarse granularity, i.e., the whole column of 128 weights, can lead to accuracy degradation.
To maintain high accuracy, 
we propose to divide the crossbar into smaller logical sub-arrays and 
only enforce the property within the fine-grained sub-array columns, and use fine-grained computation instead of coarse-grained computation (i.e., calculate fine-grained sub-array column at each time).
However, this raises another problem: 
the mainstream designs take advantage of coarse-grained computation to achieve high performance (i.e., frames per second (FPS)).
For a fine-grained design to reach a similar FPS to that of coarse-grained designs, if no other optimizations are applied, more parallel analog-to-digital converters (ADCs) are needed to operate in parallel to compensate for the performance degradation caused by computation granularity, which will generally lead to a higher hardware overhead than the coarse-grained designs.

Is this yet another failed idea that seems to work at the first thought but actually not after deliberation?
Fortunately, the answer is {\em no}, because our idea opens a new 
opportunity for more significant performance improvements.
The crucial observation is that if {\em all} inputs to the crossbar
become zeros in higher bits after a certain position, e.g., {\bf 0000}1011,
they can be simply skipped, saving cycles and improving performance. 
The effectiveness of the input zero-skipping technique is 
determined by the size of the crossbar sub-array. The smaller the number of 
inputs, the higher the probability all inputs become zero at 
the higher bits, thereby more zeros can be skipped.
This is why zero-skipping is a {\em unique} opportunity for 
small sub-arrays.
With a fraction of a crossbar column (e.g., 4 or 8 cells),
zeros can be aggressively skipped among 4 or 8 inputs. 
Moreover, it is worth noting that small ADCs for fine-grained computation are easier to implement than large ADCs for coarse-grained computation, and make the design less susceptible to non-idealities and noise~\cite{johns2008analog,kester2006adc}.

Putting all together, this paper proposes {\em FORMS},
the first algorithm/hardware co-designed fine-grained ReRAM architecture solution leveraging polarized weights.
The overall flow of FORMS algorithm/hardware co-design is shown in Figure~\ref{fig:overall_flow}.
Starting with a pretrained model, we first apply
structured pruning.
Then, the weight matrix in the pruned model is divided into fixed-sized {\em fragments}, and each fragment corresponds to a column of the crossbar sub-array.
By incorporating our fragment polarization constraints in ADMM regularized training, the weights in each fragment will be trained to have the same sign.
Please note that we are \textbf{not} moving the weights around to form polarized fragments.
Finally, the quantization is applied to reduce the number of bits
required for each weight.
With the multi-step ADMM-based structured pruning, polarization, and 
quantization, a significant model compression ratio is achieved. 

At the hardware level, 
we design a fine-grained DNN accelerator architecture leveraging fine-grained computations. 
A novel zero-skipping logic is developed to control the shift of input bit streams on-the-fly and avoids entering the unnecessary zero bits into each layer of the DNNs-based ReRAMs to eliminate useless computations and start computation on the next input early. 
Applied to the inputs of a small sub-array (fragment), zero-skipping
significantly reduces the frame processing time 
and energy consumption. 

%To make a fair ISO-area comparison, we compare the throughput and frame processing rate of FORMS to a representative ISAAC under similar power and area.
%When consuming almost the same  power and area compared to ISAAC, FORMS achieves 1.50$\times$ and 1.93$\times$ throughput improvements in terms of $\frac {GOPs}{s \times mm^{2}}$ and $\frac {GOPs}{W}$ , and 1.12$\times$ $\sim$ 2.1$\times$ speedups in terms of frame per second over ISAAC. Interestingly, FORMS optimization framework can even speed up the original unmodified ISAAC from 14.97$\times$ in the case of ResNet18 with ImageNet up to 401.64$\times$ in the case of ResNet18 with CIFAR-10, reflecting the importance of algorithm and hardware co-design.

\section{Background and Challenges}
\label{background}

\subsection{ReRAM Crossbar and ReRAM-Based DNN Acceleration}

Recently, there is significant progress in fabricating non-volatile memories. The 3D Xpoint~\cite{izraelevitz2019basic, hady2017platform} is an example of commercial non-volatile memories fabricated jointly by Micron and Intel. 
Resistive RAM is a non-volatile memory with nearly zero leakage power, high integration density, and high scalability.
Research papers demonstrated the results of fabricated ReRAM memory cells, memory arrays \cite{trinh2018resistive} and also neuromorphic accelerators using ReRAM technology~\cite{kataeva2019towards, izraelevitz2019basic}.

Several ReRAM-based in-situ mixed-signal DNN accelerators such as ISAAC~\cite{shafiee2016isaac},
Newton~\cite{nag2018newton},
PipeLayer~\cite{song2017pipelayer},
PRIME~\cite{PRIME},
PUMA~\cite{ankit2019puma},
MultiScale~\cite{marinella2018multiscale},
XNOR-RRAM~\cite{sun2018xnor},
RapidDNN~\cite{imani2018rapidnn}, have been proposed in recent years. 
These designs utilize a combination of analog and digital units to speed up the computation.
Besides, the recent work TIMELY~\cite{li2020timely} is proposed to enhance the analog data locality to keep computations in analog domain to save the energy cost of data movements and D/A and A/D domain conversion. The SRE~\cite{yang2019sparse} exploits the sparsity of both weights and activations to achieve better energy efficiency and performance by proposing hardware mechanisms. However, it induces remarkable hardware overhead as all mechanisms such as row indexing, routing controls, and word-line controls are hardware-managed. TinyADC~\cite{yuan2021tinyADC} proposes a pruning solution that fixes the number of non-zero weights in each column of the ReRAM crossbar while their positions can vary. 
This helps to decrease the accumulated value and required ADC resolution. 
In contrast, FORMS proposes hardware-software co-design optimizations to improve frame processing rate, area, and power efficiency.
%proposes to use the fine-grained operation units to improve computational accuracy. 
% The SME~\cite{liu2021sme} proposes a ReRAM-based sparse multiplication engine. To better leverage the bit-sparsity, they slice the bits of weight and map them across the crossbars, then combine the activation results using peripheral circuits.

\subsection{Challenges}
\label{sec:challenge}

Despite the recent research progress, we identify two challenges
for ReRAM-based DNN acceleration. 

{\bf Mapping signed weights. }
When we map the arbitrary DNN weights onto ReRAM crossbars, 
it is challenging to represent negative weights 
using positive conductance. 
Prior works address the problem differently. 
The general way is to decompose each weight into a positive magnitude portion and a negative magnitude portion, then use two crossbars to represent each of the portions~\cite{li2018efficient,liu2020neural, krestinskaya2020memristive,sayyaparaju2017circuit,PRIME}.
This method doubles the hardware cost of the ReRAM crossbar. 
Instead, ISAAC~\cite{shafiee2016isaac} addresses this problem by shifting or adding an offset to the original negative weight value, so that all the weights become positive. 
To be able to calculate correct results, ISAAC needs to count the number of 1’s in each individual input. If considering 16-bit weights are used, for each 1 in each input, a bias of $2^{15}$ must be subtracted from the final results. Counting all 1s for all inputs (that feed to the crossbar in parallel) and performing subtractions for each of 1’s introduces significant overhead to ISAAC. Moreover, this mapping method decreases the network robustness to hardware failures~\cite{yuan2021failure}.
We can see that both methods will cost extra resources in terms of area, power, and energy consumption.

{\bf ADC/DAC implementation}. 
To ensure high throughput, the mainstream accelerator designs (e.g., ISAAC, PRIME, PUMA) take advantage of coarse-grained computations, a large number of ReRAM crossbar rows (e.g., 128 or 256) need to be processed at the same time and hence large ADCs are needed.
However, the ADC does not scale as fast as the CMOS technology does~\cite{imani2019floatpim, shafiee2016isaac}.
The recent study~\cite{imani2019floatpim, yuan2021tinyADC} reports the ADC/DAC blocks may become the major contributor to the total chip area and power.
To save power and area, many designs share an ADC with many crossbar columns (e.g., 128 columns in ISAAC). 
Therefore, the ADC needs to switch among those columns at a high-speed, which is hard in practice.
% in one cycle. 
% In practice, it is hard to fabricate a chip that its ADC can switch between many columns too fast.

\subsection{Motivation}
\label{sec:motivation}
% Our research is motivated by the two challenges. 
We realize the possibility of generating polarized weights that can elegantly solve the natural problem of mapping positive/negative weights to the ReRAM crossbar without doubling crossbar cost or introducing extra hardware for result compensation.
We explore a good balance between the overall hardware cost and performance. 
Even with paying the extra overhead to compensate for the performance degradation caused by fine-grained computation, our design can still achieve four benefits:
1) zero-skipping techniques can achieve significant performance 
improvements; 
2) making ADC/DAC implementation less challenging compared to the 
state-of-the-art coarse-grained architecture designs (e.g., ISAAC, PUMA, and PRIME);
3) fine-grained architecture is less susceptible to non-idealities and noise than coarse-grained architecture;
and 4) achieving a higher overall performance rate than coarse-grained designs under a similar power and area cost.
We demonstrate that by leveraging the principle of 
algorithm and hardware co-design, our proposed solution 
significantly advances the state-of-the-art and paves a new way for fine-grained mixed-signal accelerators design.

\section{Hardware-Aware Optimization \\ Framework}

In this section, we describe the whole software procedure
to generate hardware-friendly DNN models. The key novelty
is the {\em fragment polarization} technique that enforces the 
same sign for weights in each fragment. As recent works~\cite{cai2020yolobile,ma2019resnet,ma2021non,li2020ss}
have demonstrated, the structured pruning and quantization 
are two essential steps for hardware-friendly model compression that
are universally applicable to all DNN accelerators. Thus, we perform
structured pruning before fragment polarization considering
the size of the ReRAM crossbars, and quantization after.
Thus, the sign of each fragment is determined by the 
structurally pruned model. 
And the quantization is
applied after the structure and sign of the weights are determined.
All of the three steps are uniformly
supported by Alternating Direction Method of Multipliers (ADMM) regularized optimization method~\cite{boyd2011distributed} which has shown to 
be very suitable for DNN training and produced state-of-the-art 
results~\cite{tiny_but_accurate}. In the following sections, 
we describe these three steps, and then explain how to express
our desired pattern in ADMM.

\begin{figure} [t]
     \centering
%      \vspace{-1.2em}
     \includegraphics[width=0.9\columnwidth]{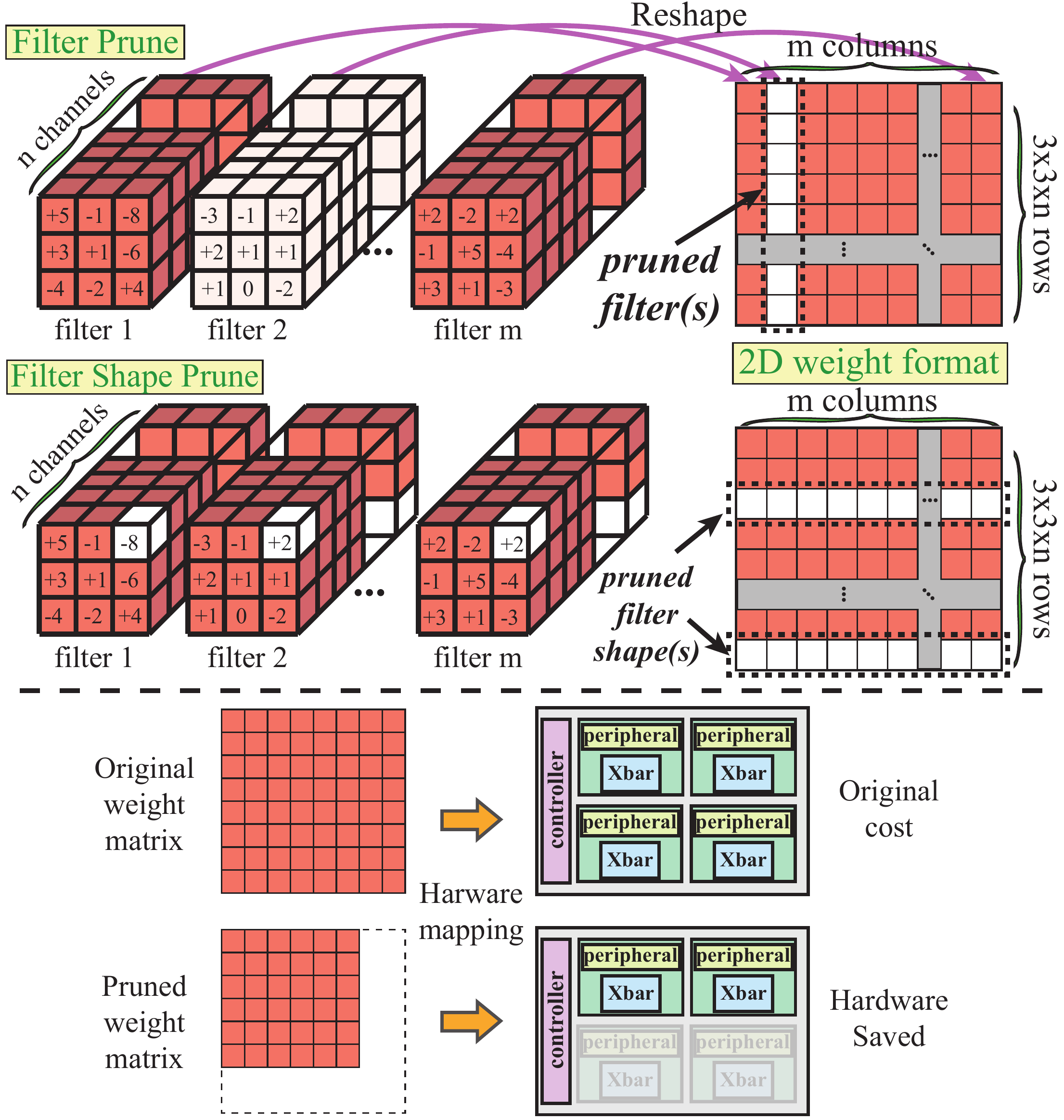}  
     \caption{Structured pruning}
     \label{fig:structured_pruning_HPCA2}
\end{figure}

\subsection{Crossbar-Aware Structured Pruning} 
There are different types of structured sparsity in DNNs.
FORMS combines two types of structured pruning methods---filter pruning and filter-shape pruning.
As shown in Figure~\ref{fig:structured_pruning_HPCA2}, we reshape the weights from convolutional filters of a convolutional layer into a 2D weight matrix, each column represents all the weights from the same filter, while each row represents the weights on the same position of all filters. By adopting these two types of pruning, the entire columns and rows of the weight matrix will be removed while the remaining weight matrix is still dense.
After the structured pruning, the weight matrix size becomes much smaller, which effectively reduces the number of ReRAM crossbars to store the weights and also removes the corresponding peripheral circuits.

The previous ReRAM-based accelerator designs~\cite{yuan2019Ultra, tiny_but_accurate} apply structured pruning and aim to make the pruning ratio as high as possible while maintaining an acceptable accuracy loss. They do not take the ReRAM crossbar size into consideration, therefore causing performance degradation when the pruned model is mapped onto the ReRAM crossbars.
For example, if the size of the crossbar is 128$\times$128, only the portion of pruned columns/rows of weights that reaches the multiple of the 128 (e.g., 128, 256) will lead to the actual crossbar reduction.
The remaining pruned columns/rows of weights still need to be stored as zeros on the crossbars. Consequently, those pruned columns/rows are wasted, and the accuracy drop is incurred without gaining the full benefit. 
In FORMS, we perform a \emph{crossbar-aware structured pruning} by considering the crossbar size and carefully choosing the pruning ratio for each DNN layer to avoid unnecessary accuracy drop.

\subsection{Fragment Polarization}
\begin{figure} [t]
     \centering
%      \vspace{-1.2em}
     \includegraphics[width=1.0\columnwidth]{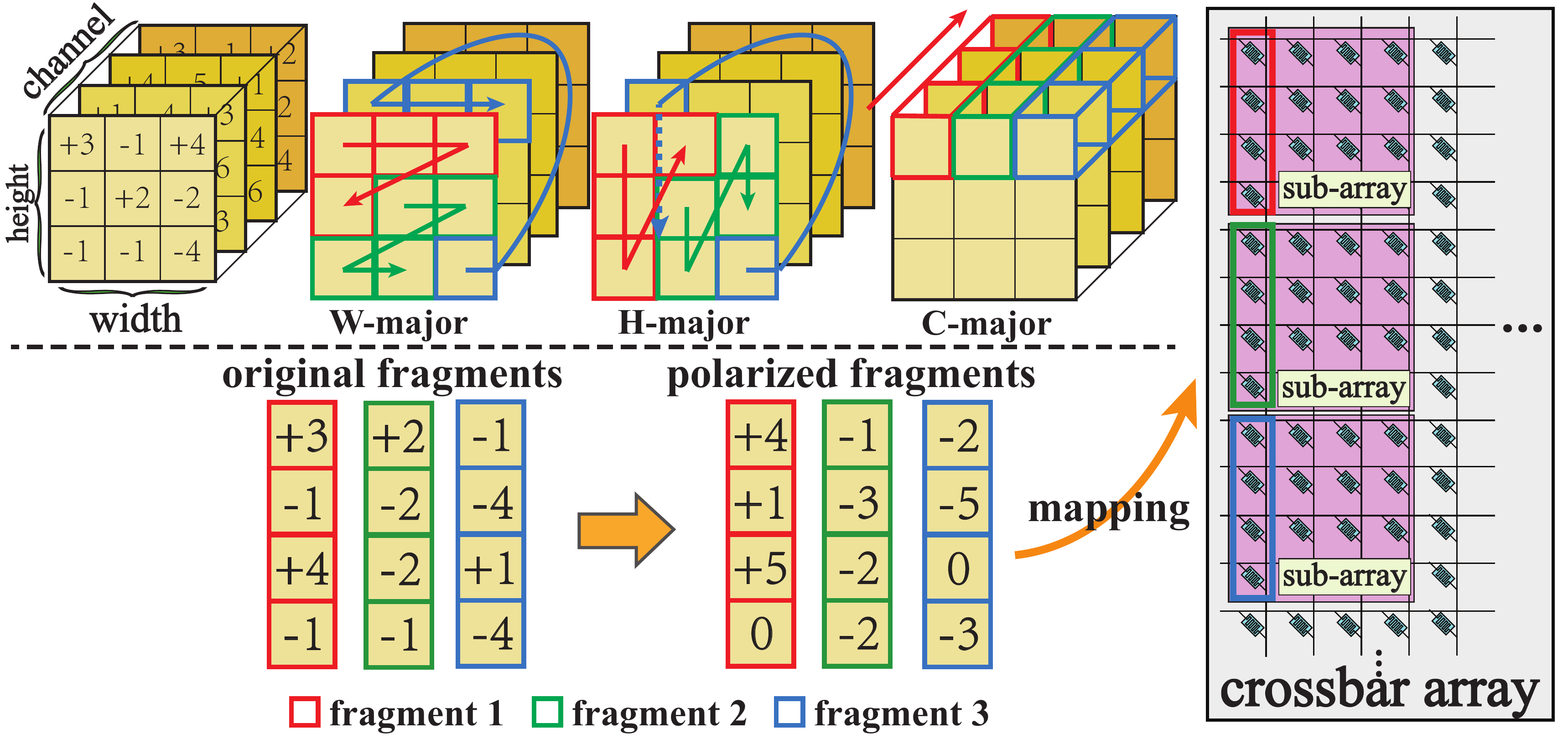} 
     \caption{Fragments and direction of polarization}
     \label{fig:polar_direction}
\end{figure}

A {\em fragment} is a set of consecutive weights
mapped to ReRAM crossbar.
The fragment size is determined by the number of rows in sub-array---a fragment contains weights
that will be mapped to the same column of a sub-array, as shown in Figure~\ref{fig:polar_direction}.
In our design, the weights in a fragment are polarized. 
They are either positive or negative. 
Therefore, the multiply-accumulate operations performed by each column of the ReRAM crossbar sub-array do not suffer from the arbitrary sign of its operands.

With the concept of fragment introduced,
we need to consider two issues:
1) how to determine the sign of weights in a fragment; and 2) the mapping policy of weights to sub-array
columns. 
The second problem matters because it determines
which set of weights in the DNN model will contain
the same sign. Next, we show how to handle these
two problems. 

To orchestrate fragment polarization during the training process, we calculate the sum of weights in each fragment and determine the fragment sign based on the following principle: if the sum is greater than or equal to 0, we set the sign of the fragment as \textit{positive}, otherwise, we set it as \textit{negative}. We incorporate ADMM regularized optimization with the fragment polarization constraint to regularize the weights to have the same sign as the fragment or to become zero. Eventually, the negative/positive weights are eliminated in positive/negative fragments.

During training, since the weights are 
continuously updated, the sign of a fragment changes. Specifically, at the beginning of fragment polarization,
the fragment signs are determined by the above 
policy based on the structurally pruned model. 
Then, the training for fragment polarization starts.
Let us assume that 
the training
process contains $N$ epochs,
each goes through the entire training dataset once. 
We can update the target sign every $M$ epochs. 
This means that at the end of $M$ epochs, 
the sign for each fragment will be calculated 
again using the {\em current} weights belonging
of the fragment. 
During the training process,
the fragment signs are updated for $N/M$ times.

Next, we consider the design space of mapping weights
to fragments. 
As shown in Figure~\ref{fig:polar_direction}, given the weights of a convolution filter in 3-D format with width (W), height (H), and
channel (C) dimension, we can naturally have three
polarization policy. 
In width-major or W-major polarization, 
the consecutive weights in one or multiple 
consecutive rows in a filter
are mapped to the same fragment.
The weights in two consecutive fragments
are also consecutive in the width-major order. 
After all weights in a filter are mapped,
we move on to the next filter using the same procedure.
More details regarding the ReRAM mapping scheme will be discussed in Section~\ref{sec:FORMS}.

Similarly, we can define height-major or 
H-major polarization, and channel-major or C-major polarization.  
In particular, for C-major, the weights in the same position at all channels are mapped before moving to the next position of the filter. 

Since the polarization policy determines which region of weights 
should have the same sign, different policies could affect 
accuracy. 
Our observation results show the W-major polarization scheme achieves the highest accuracy on the ImageNet dataset, where $C-major$ polarization is the best choice for the CIFAR-10/100 dataset.
Since the same polarization scheme will be applied over the entire neural network, we only need to uniformly re-order the weights with their corresponding inputs {\em in advance}, then directly map them to the ReRAM crossbars.
Base on the chosen polarization scheme, the weights are trained to enforce the fragment polarization properties. There is no need to individually move the weights around to form a polarized fragment, and it will not incur any hardware overheads due to location indices.

\subsection{ReRAM Customized Weight Quantization}
We develop ReRAM customized weight quantization, to 
reduce the bit representation of weight parameters. Similar to crossbar-aware structured pruning, we consider the numbers of valid ReRAM conductance states into our quantization constraint during the training process.
In reality, the number of ReRAM conductance states is limited by the provided resolution of the peripheral write and read circuitry. 
More state levels require more sophisticated peripheral circuitry. 
Due to limited computational accuracy, multiple ReRAM cells are usually used to represent one weight. 
For example, we need eight 2-bit ReRAM cells to represent one 16-bit weight, and four 2-bit ReRAM cells to represent one 8-bit weight. 
Thus, the quantization can effectively reduce the design area and power consumption.
In FORMS, if the 2-bit ReRAM cells are used in the design, the quantization bits will be set to the multiply of 2, which can fully utilize the resolution of the ReRAM cells.
More importantly, compared to forcing a high-bit representation DNN model into a low-bit representation during the ReRAM mapping process, introducing the quantization to the training process allows the weights to be trained into a low-bit representation.

\subsection{ADMM-Regularized Optimizations}
\label{sec:admm}

ADMM~\cite{boyd2011distributed} is an advanced optimization technique, where an original optimization problem is decomposed into two sub-problems that can be solved separately and iteratively. We incorporate ADMM-regularized training into FORMS training process to achieve and optimize the crossbar-aware structured pruning, fragment polarization, and ReRAM customized quantization feature. Using ADMM can guarantee the solution feasibility (satisfying ReRAM hardware constraints) while providing high solution quality (no obvious accuracy degradation after model compression and after hardware mapping).

Consider the $i$-th layer in an $N$-layer DNN, the weights and bias can be represented by ${\bf{W}}_{i}$ and ${\bf{b}}_{i}$. And we define the loss function as: $\mathcal{L} \big( \{{\bf{W}}_{i}\}_{i=1}^N, \{{\bf{b}}_{i} \}_{i=1}^N \big)$. We minimize the loss function associated with DNN model and subject to the constraints of FORMS. Then the overall problem is given by:
\vspace{-0.3em}
\begin{equation}
\small
\label{equ1}
\begin{aligned}
& \underset{ \{{\bf{W}}_{i}\},\{{\bf{b}}_{i} \}}{\text{minimize}}
& & \mathcal{L} \big( \{{\bf{W}}_{i}\}_{i=1}^N, \{{\bf{b}}_{i} \}_{i=1}^N \big),
\\ & \text{subject to}
& & {\bf{W}}_{i}\in {\bf{S}}_{i}, \; {\bf{W}}_{i}\in {\bf{P}}_{i}, \; {\bf{W}}_{i}\in {\bf{Q}}_{i}, \; i = 1, \ldots, N,
% & & {\bf{W}}_{i}\in {\bf{S}}_{i}, \; i = 1, \ldots, N.
\end{aligned}
\end{equation}
where ${\bf{S}}_{i}$, ${\bf{P}}_{i}$ and ${\bf{Q}}_{i}$ are the constraint sets of Crossbar-aware Structured Pruning, Fragment Polarization and ReRAM Customized Quantization, respectively. 

\subsubsection{\textbf{Crossbar-aware Structured Pruning Constraints}}
In crossbar-aware structured pruning, we use $\bf{H}$ to represent the weights in 2D format of a specific layer, as shown in Figure~\ref{fig:structured_pruning_HPCA2}. 
The constraints in the $i$-th CONV layer becomes
${\bf{W}}_{i}\in {\bf{S}}_{i}
:=
\{{\bf{H}}\mid$ {the percentage of nonzero \textit{filters} and \textit{filter-shapes} in} ${\bf{H}}$ {is less than or equal to} $\alpha_i$ and $\beta_i$, 
where $\alpha_i$ and $\beta_i$ are predefined hyperparameters. For example, suppose we want a 43\% filter sparsity and 62\% shape sparsity in $i_{th}$ layer, then we set $\alpha_i=0.57$ and $\beta_i=0.38$.

\subsubsection{\textbf{Fragment Polarization Constraints}}
For fragment polarization, the constraint set ${\bf{P}}_{i}$ = \{the weights on each fragment (a column of a crossbar sub-array) have the same sign\}, where the ReRAM crossbar sub-array size $m \times n$ is a predefined hyperparameter. The sign of the fragment is determined by the following function:
\vspace{-0.5em}
\begin{equation}
\label{equ2}
    {Sign}_{\bf{f}}=
\begin{cases}
     + & \text { if } \sum_{i=1}^{m}({{W}}_{i}) \ge 0, \\ 
 - & \text { otherwise, }
\end{cases}
\end{equation}
% \vspace{-0.8em}
where $i \in 1, \ldots, m$ is the $i$-th weight on a fragment.

\subsubsection{\textbf{Customized Quantization Constraints}}
For the ReRAM customized quantization, the set ${\bf{Q}}_{i}$ $=\{$the weights in
the $i$-th layer are mapped to the quantization values$\}$. The quantization values depend on the characteristics of the ReRAM device such as conductance range and valid state levels.

\subsubsection{\textbf{ADMM-Regularized Optimization Flow}}
\begin{figure} [t]
     \centering
%      \vspace{-1.2em}
     \includegraphics[width=0.8\columnwidth]{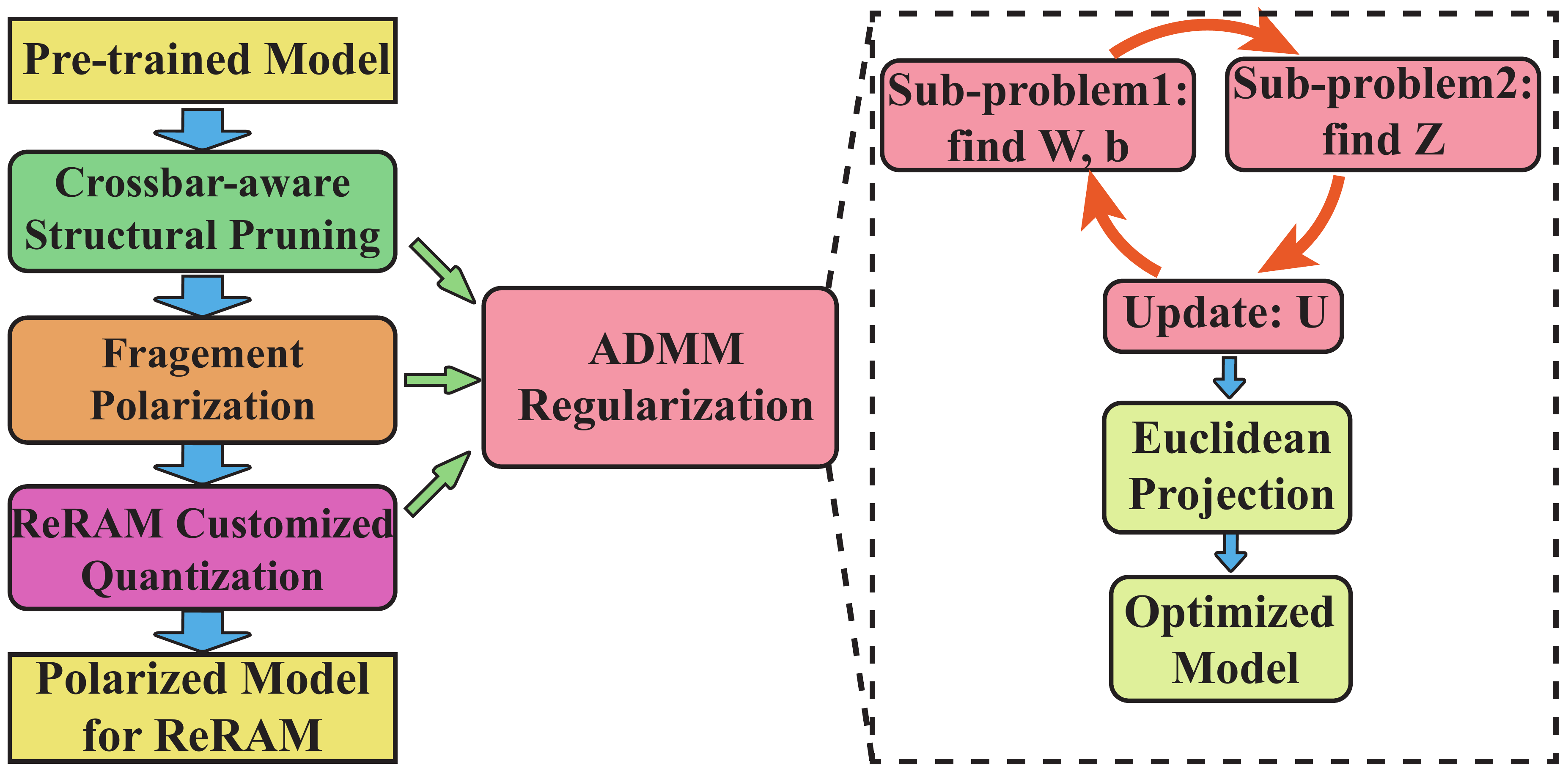}     \caption{Procedure of ADMM-Regularized Optimization.}
     \label{fig:procedure}
\end{figure}

As shown in Figure~\ref{fig:procedure}, the overall ADMM-regularized optimization flow is an iterative training process, which is similar in ADMM-NN~\cite{admm_nn}.
Thanks to the flexibility in the definition of constraint sets ${\bf{S}}_i$, ${\bf{P}}_i$, and ${\bf{Q}}_i$, the above constraints can be jointly included, or applied individually. 
First, we incorporate one of the three constraints by using indicator function, which is

$
% \mathcolorbox{yellow}{
g_{i}({\bf{W}}_{i})=
\begin{cases}
 0 & \text { if } {\bf{W}}_{i}\in {\bf{S}}_{i}\ or\ {\bf{P}}_{i}\ or\ {\bf{Q}}_{i}, \\ 
 +\infty & \text { otherwise.}
\end{cases}
$

Problem~(\ref{equ1}) with constraint cannot be directly solved by classic stochastic gradient descent (SGD) methods~\cite{kingma2014adam} as original DNN training.
However, the ADMM regularization can reforge and separate the problem, then solve them iteratively~\cite{hong2016convergence,liu2018zeroth}. First, we reformulate problem (\ref{equ1}) as follows:
\vspace{-1em}
\begin{equation}
\label{equ3}
\begin{aligned}
% \mathcolorbox{yellow}{
& \underset{ \{{\bf{W}}_{i}\},\{{\bf{b}}_{i} \}}{\text{minimize}}
& & f \big( \{{\bf{W}}_{i} \}_{i=1}^N, \{{\bf{b}}_{i} \}_{i=1}^N \big)+\sum_{i=1}^{N} g_{i}({\bf{Z}}_{i}),
\\ & \text{subject to}
& & {\bf{W}}_{i}={\bf{Z}}_{i}, \; i = 1, \ldots, N,
% }
\end{aligned}
\end{equation}
\vspace{-0.2em}
where ${\bf Z}_{i}$ is an auxiliary variable.
With formation of augmented Lagrangian~\cite{boyd2011}, problem (\ref{equ3}) can be decomposed into two subproblems \eqref{subproblem_1} and \eqref{subproblem_2},
\vspace{-0.5em}
\begin{equation}
\small
\label{subproblem_1}
 \underset{ \{{\bf{W}}_{i}\},\{{\bf{b}}_{i} \}}{\text{minimize}}
\ \ \ f \big( \{{\bf{W}}_{i} \}_{i=1}^N, \{{\bf{b}}_{i} \}_{i=1}^N \big)+\sum_{i=1}^{N} \frac{\rho_{i}}{2}  \| {\bf{W}}_{i}-{\bf{Z}}_{i}^{t}+{\bf{U}}_{i}^{t} \|_{F}^{2}, \\
\end{equation}
\vspace{-0.4em}
\begin{equation}
\small
\label{subproblem_2}
\underset{ \{{\bf{Z}}_{i} \}}{\text{minimize}}
\ \ \ \sum_{i=1}^{N} g_{i}({\bf{Z}}_{i})+\sum_{i=1}^{N} \frac{\rho_{i}}{2} \| {\bf{W}}_{i}^{t+1}-{\bf{Z}}_{i}+{\bf{U}}_{i}^{t} \|_{F}^{2}, \\
\end{equation}
% \vspace{-0.1em}
where ${\bf U}_{i}$ denotes dual variable and $t$ is the iteration index. The positive scalars $ \rho_{i}$ is a penalty hyperparameter for the $L2$ regularization. 
The first subproblem can be solved by classic SGD, and the solution for the second subproblem is given by
     \vspace{-0.5em}
\begin{equation}
\small
{\bf{Z}}_{i}^{t+1} = \prod_{{\bf{X}}_{i}}({\bf{W}}_{i}^{t+1}+{\bf{U}}_{i}^{t}),
     \vspace{-0.5em}
\end{equation}
     \vspace{-0.2em}
where $\prod_{{\bf{X}}_{i}}(\star) $ is Euclidean projection to ${{\bf{X}}_{i}}\in\{{\bf{S}_i}, {\bf{P}_i}, {\bf{Q}_i}\}$, thereby weight matrices are structured pruned or fragment polarized or customized quantized. These two subproblems will be iteratively solved and we update ${\bf U}_{i}$ in each iteration by ${\bf{U}}_{i}^{t}:={\bf{U}}_{i}^{t-1}+{\bf{W}}_{i}^{t}-{\bf{Z}}_{i}^{t}$ until convergence. 
The detailed solution process can refer to~\cite{admm_tianyun}.

\section{FORMS Architecture}

This section describes the FORMS accelerator architecture that can 
execute the optimized DNN models generated by the 
FORMS optimization framework.
Our design only needs to map the magnitude bits of all the weights to ReRAM crossbars without adding extra crossbars or offset circuits.
To ensure high throughput, 
we develop a new pipelined design incorporated with zero skipping logic.

\begin{figure} [b]
     \centering
%      \vspace{-1.2em}
     \includegraphics[width=0.9\columnwidth]{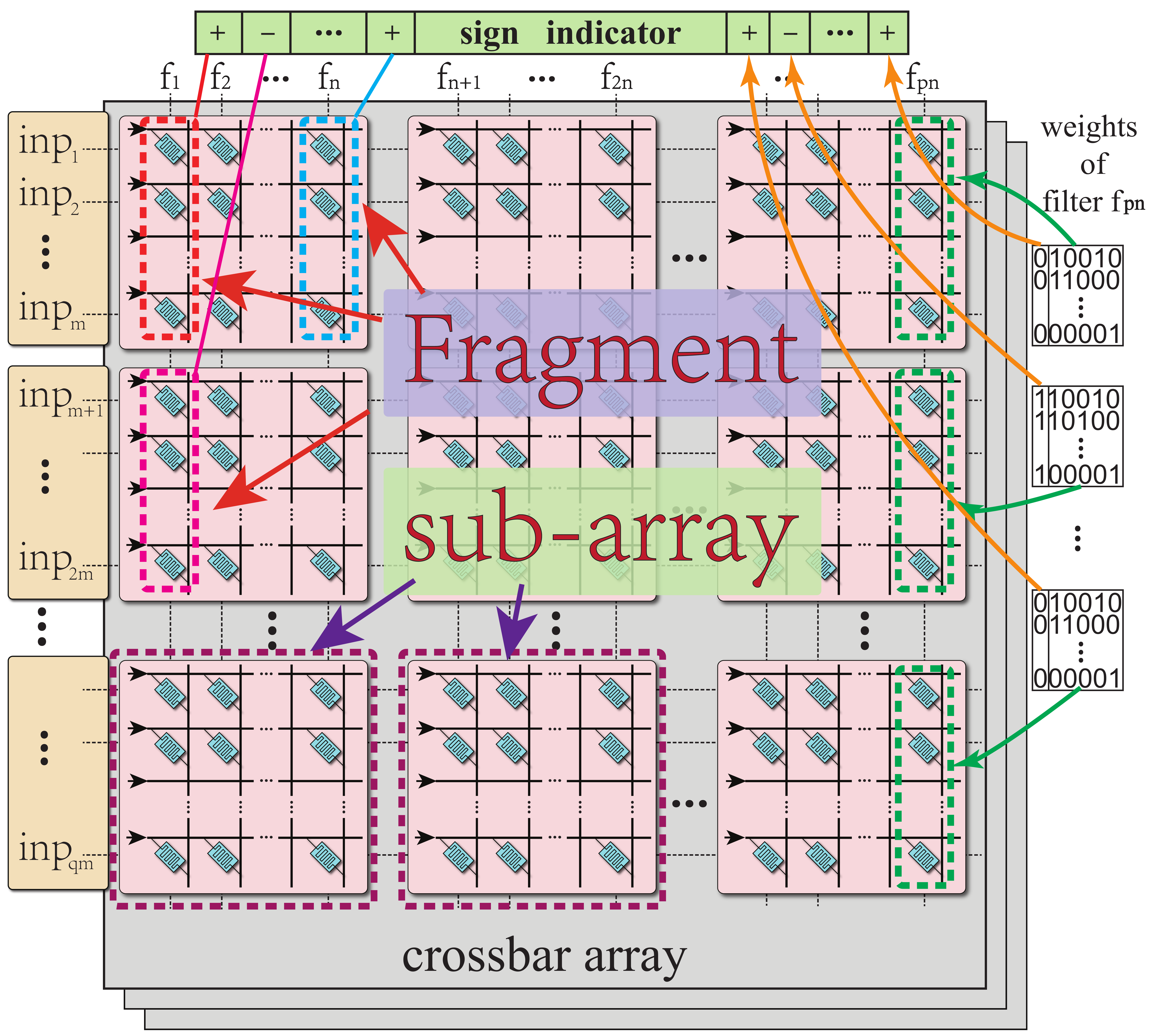}     \caption{Polarized ReRAM Crossbar Mapping Scheme.}
     \label{fig:polarized}
\end{figure}

\subsection{Mapping Scheme and Dataflow}
\label{sec:FORMS}

In FORMS, instead of mapping both the sign and magnitude bits of arbitrary/mixed signs of weight values as in prior works such as memristive Boltzman machine~\cite{Boltzman_Machine}, ISAAC~\cite{shafiee2016isaac}, Newton~\cite{nag2018newton}, PipleLayer~\cite{song2017pipelayer}, PRIME~\cite{PRIME}, PUMA~\cite{ankit2019puma}, we only need to store the magnitude bits to the ReRAM, and the sign bit of each fragment in the 1R ReRAM-based array~\cite{chen2003access}. 

Figure~\ref{fig:polarized} shows an example of the ReRAM mapping scheme on a crossbar array.
We assume the physical crossbar array size is ${q}\times{m}$ rows by ${p}\times{n}$ columns (e.g., 128 by 128), where $q, p, m, n$ could be various depending on different design specifications. 
We partition the crossbar array into logical sub-arrays, where each sub-array has $m$ rows and $n$ columns.
On one crossbar array, ${q}\times{m}$ weights from the same filter will be mapped on the same column, and the weights from different ${p}\times{n}$ filters will be mapped onto different columns of this crossbar array.
The rest of the weights will be mapped onto other crossbar arrays in the same manner. 
Thus, in order to accommodate all the weights of a CONV layer, multiple ReRAM crossbar arrays are needed. In reality, due to the limitation of the ReRAM resolution, we need multiple ReRAM cells to represent one weight. For example, we need four 2-bit ReRAM cells to represent one 8-bit weight. In this way, each fragment will still have $m$ rows, but 4 columns instead of 1 column. And all $m$ weights in the same fragment that are represented by those $m\times 4$ ReRAM cells.

The optimized ReRAM-aware DNN model obtained from the ADMM regularized optimization process in Section~\ref{sec:admm}, 
has the fragment polarization property.
Thus, all the weights within a fragment have the same sign, and we do not need to move the weights around to satisfy the polarization constraint.
For each fragment, an associated {1R} is used to store the sign bit. All the associated {1Rs} are grouped as a sign indicator. The sign bits stored in the sign indicator will be carried out for accumulation in the digital domain.

\noindent \textbf{Convolution Dataflow.}
CONV layers perform convolution of input feature maps and weight filters.
The convolution results are then accumulated. After passing intermediate results through an activation function (e.g., rectified linear unit (ReLU)), we produce a single output feature map. By repeating this procedure for the rest of the weight filters, we obtain the whole output feature maps. 

We fetch the digital inputs (feature maps) from eDRAM or DRAM (DRAM for input images, eDRAM for intermediate results) and arrange them into input buffers to be sent to DAC. The output of DAC becomes the analog input $\mathbf{v}_i$ of the ReRAM crossbars. The matrix-vector multiplication can be performed by leveraging the feature of the ReRAM crossbars, and the output can be calculated by: $\textbf{i}_o=\mathbf{W}^\mathbf{T}\mathbf{v}_i$.
Each fragment (column) of the ReRAM crossbar sub-array produces an intermediate result, which is the accumulated current result.
Intermediate results produced by all fragments will be propagated through ADC. The converted digital values will be carried to our proposed accumulation blocks  along with the corresponding sign bits from sign indicator for accumulation.

The accumulation blocks are used for accumulating all the intermediate results from crossbar sub-arrays. 
The sign-bit indicator specifies whether the adder should work in add or subtract mode. The output of the adder will be accumulated to the temporary results from other fragments.
Iteratively, we finish the CONV operation and obtain the output feature map to be stored into eDRAM, which will become the inputs (feature maps) of the next layer.
By using the sign indicator with our polarized ReRAM crossbar mapping, our design can 
save half of the crossbars, which are used to store the positive/negative weights separately.
When compared to the designs like ISAAC~\cite{shafiee2016isaac},
FORMS avoids introducing the offset circuity with small overhead caused by sign indicator.
Moreover, since we only store the magnitude bits on ReRAM crossbars, FORMS can take the advantage to store one more magnitude bit when using the same number of ReRAM crossbars, which allows FORMS to achieve higher weight representation precision.

\subsection{Fragment Size Exploration and Zero Skipping Logic}
\label{sec:zero-skip}

\begin{figure} [t]
     \centering
%      \vspace{-1.2em}
     \includegraphics[width=0.8\columnwidth]{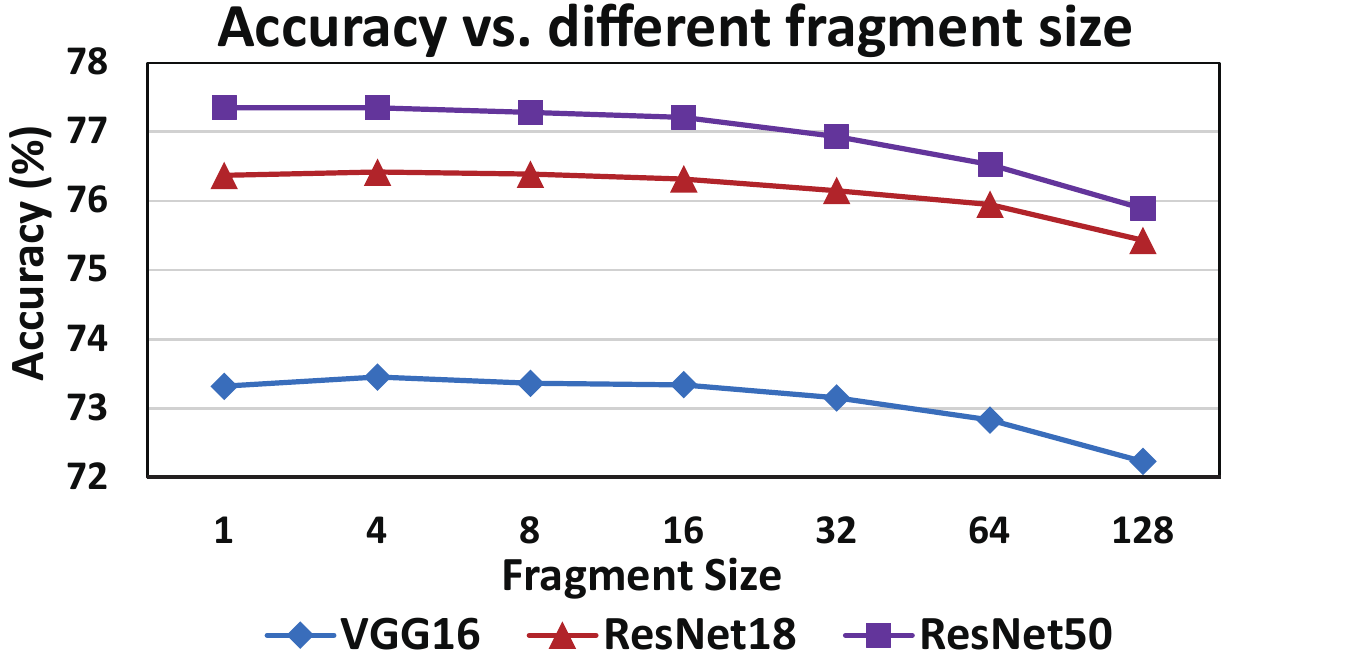}    
     \caption{Test accuracy under different fragment sizes on CIFAR-100 dataset.}
     \label{fig:acc_vs_frag_size}
 \end{figure}

The fine-grained sub-array is a key feature of FORMS architecture
that ensures high accuracy, more feasible hardware implementation, 
and significant performance improvement. 
We perform experiments to understand the relation 
between the size of fragment size and accuracy. 
Figure~\ref{fig:acc_vs_frag_size} shows that
the smaller fragment size introduces zero or minor accuracy degradation in polarized form, while the larger fragment size may lead to a small accuracy degradation.
Compared to the coarse-grained crossbars, 
the small fragment size corresponds to small ADCs. For example, FORMS can use 4-bit ADCs instead of one 8-bit ADC per each crossbar used in ISAAC~\cite{shafiee2016isaac} or PROMISE~\cite{srivastava2018promise}. 
In general, the area and power of ADCs grow exponentially 
with the number of bits of ADCs.
To save the area and power, ISAAC shares a large 8-bit ADC with 128 columns. Unfortunately, this leads to stringent hardware implementation requirement.
Specifically, one 8-bit ADC must switch between 128 columns and convert the analog current to digital values within 100ns.
This design makes ISAAC impractical for fabrication. 
More importantly, having fine-grained fragments opens
a new opportunity to significantly improve performance by 
employing a novel zero skipping logic that reduces the required cycles to feed input bits to sub-arrays, leading to significant
performance improvement.

\begin{figure} [b]
     \centering
%      \vspace{-1.2em}
     \includegraphics[width=0.6\columnwidth]{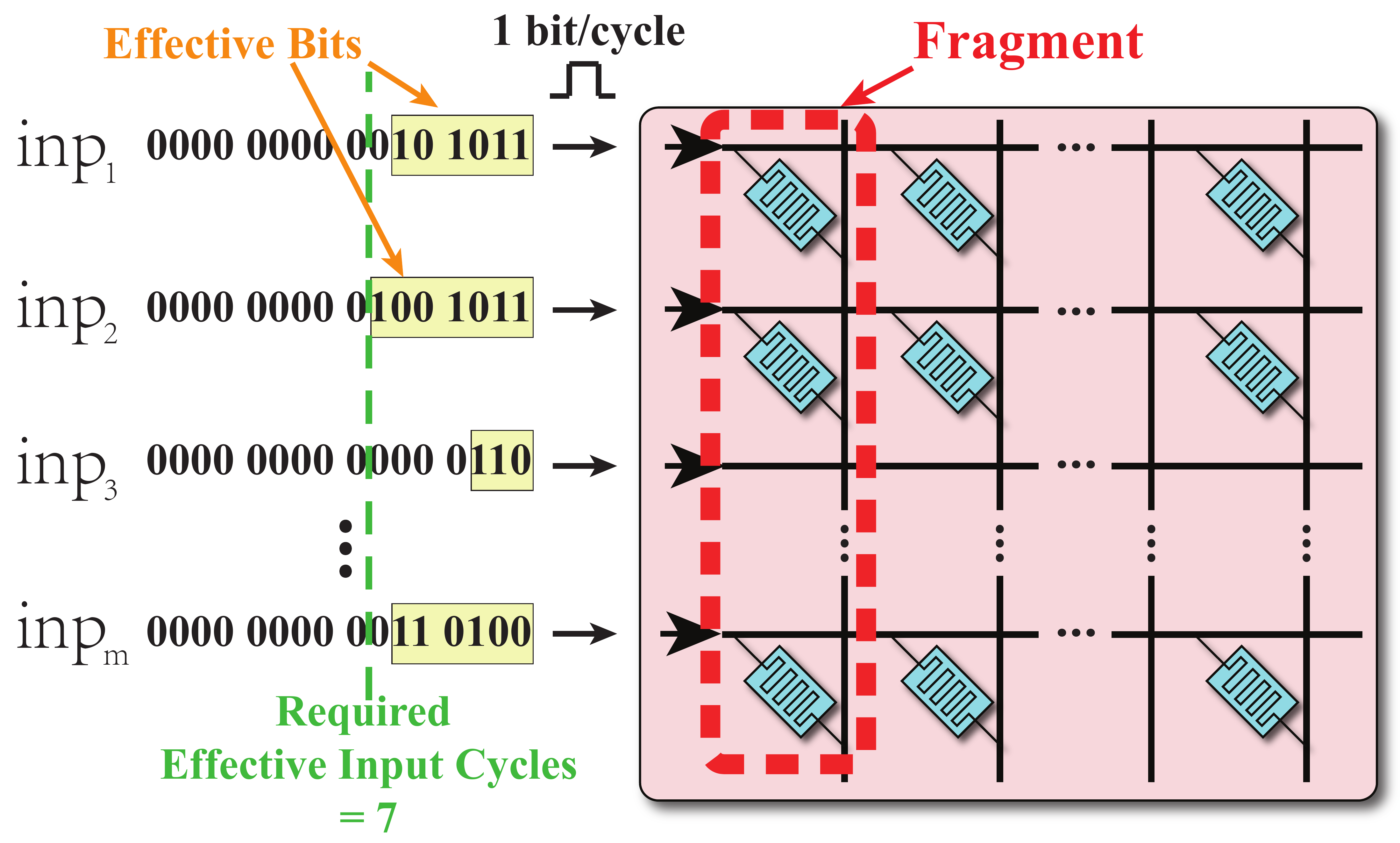}    
     \caption{Input Effective Bits and Required Fragment EIC.}
     \label{fig:effective_bits}
 \end{figure}

\begin{figure} [t]
     \centering
%      \vspace{-1.2em}
     \includegraphics[width=1\columnwidth]{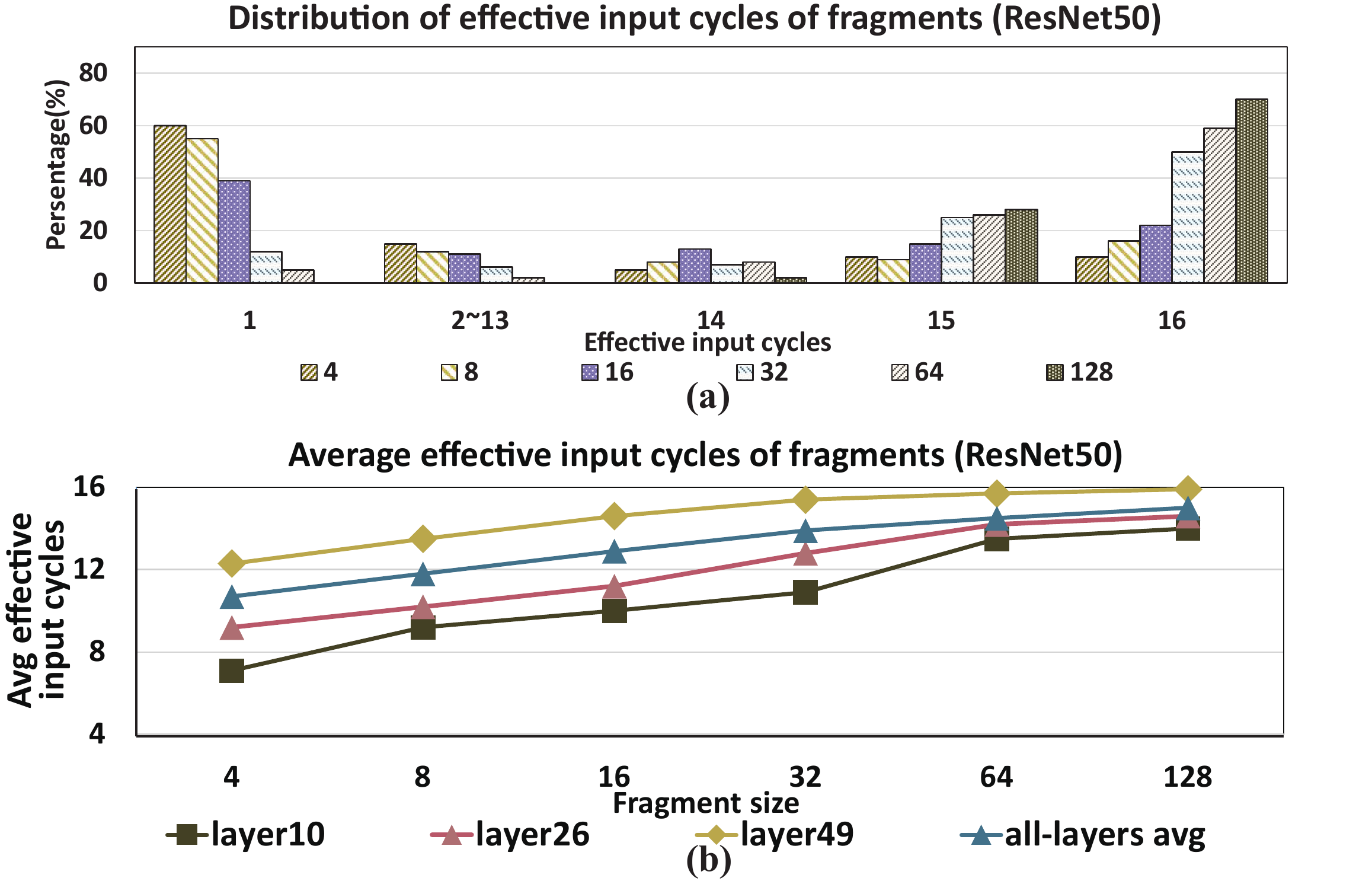}    
     \caption{(a) Percentage of effective input cycles for different fragment sizes using 16-bit input data; (b) average effective input cycles for various fragment sizes.}
     \label{fig:input_dist}
 \end{figure}

To understand how FORMS saves input cycles, we use a 16-bit input representation as an example. For each clock cycle, 1 input bit is fed into the fragments in parallel. Generally, it needs 16 cycles to feed a 16-bit input to the crossbar. However, most inputs actually have small values~\cite{batch_norm} and can be represented by just a few bits, which means the upper bits are all 0s and can be skipped from feeding to the crossbar.
% and we do not need to feed those upper bits into the crossbar.

We define {\em effective bits} as the number of input bits that contribute to output results. It is obtained by removing the consecutive most significant zeros among all inputs. We also define the {\em effective input cycles (EIC)} as the minimum number of required cycles to fed effective bits of all inputs into a fragment, which is equal to the maximum effective bits of all inputs corresponding to that fragment. For example, as shown in Figure~\ref{fig:effective_bits}, the effective bits of $inp_1$ is 6, however, the required EIC for the fragment is 7. Because the $inp_2$ has the largest effective bits in the fragment, which is 7.
The EIC in the coarse-grained designs that have large fragment size (e.g., 128) is generally higher than the fine-grained designs that have smaller fragment size (e.g., 4,8 or 16). The reason is that the larger fragments have a higher probability to contain the inputs that have higher effective bits. Figure~\ref{fig:input_dist} (a) shows an example of the percentage of EIC (1 to 16) of fragments using 16-bit inputs and various fragment sizes (i.e., 4, 8, 16, 32, 64, 128) in one CONV layer. As the fragment size becomes larger, it increases the percentage of the fragments that require more EIC. Figure~\ref{fig:input_dist} (b) shows, under different fragment sizes, the average required EIC over all fragments for different layers. 
For fragment size of 4, 
the average required EIC of all layers to 
feed all effective bits into crossbar is 10.7, which saves 33\% of the total 16 cycles.
The required EIC for fragment size of 128 is 15, which only saves 6\% cycles.
These results show that while zero-skipping is applicable to
coarse-grain crossbars as well, the benefits are drastically less.

\begin{figure} [b]
     \centering
%      \vspace{-1.2em}
     \includegraphics[width=0.7\columnwidth]{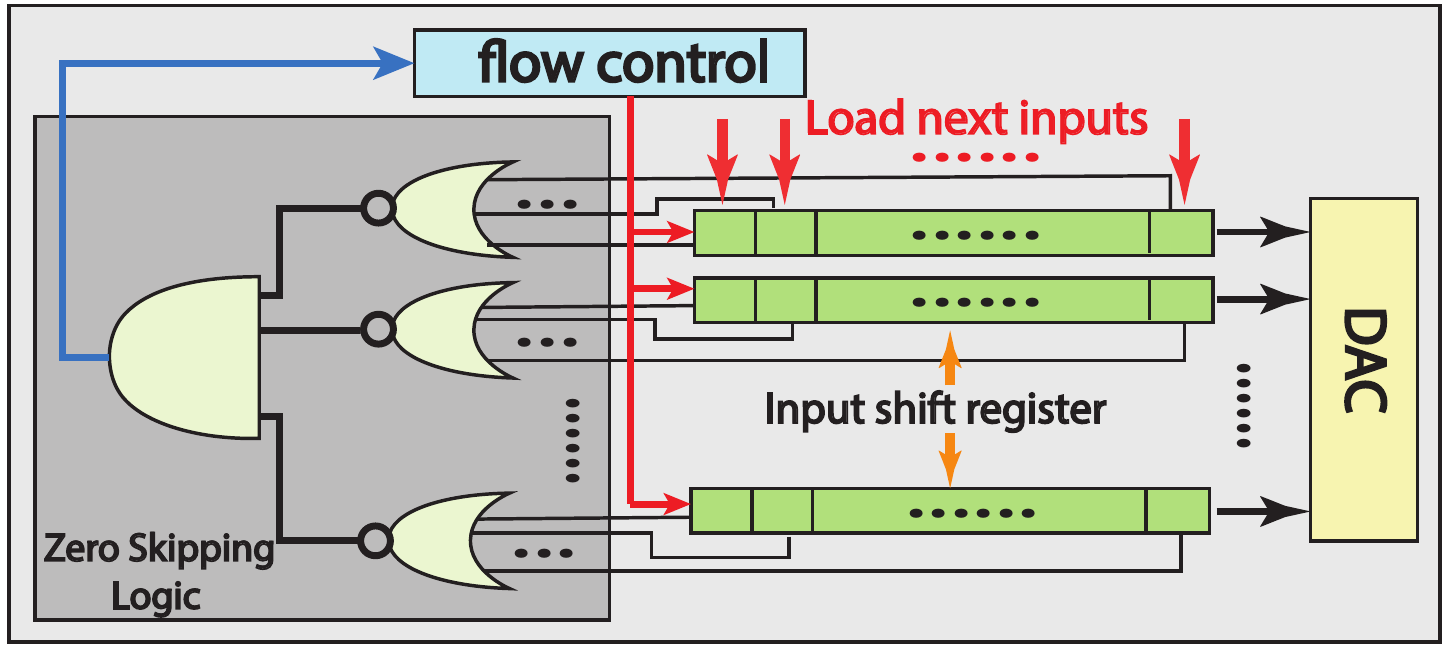}     
     \caption{Zero-skipping Logic.}
     \label{fig:zero_skip}
\end{figure}

To take this key advantage, 
we propose a zero skipping logic as shown in Figure~\ref{fig:zero_skip}.
It dynamically controls the input cycles to skip the redundant cycles for feeding zeros, which do not contribute to output, but consume power, energy and increase latency.
As a result, FORMS achieves higher
performance than the coarse-grained designs.
% by eliminates more wasted cycles.

\subsection{Overall Architecture}  
\label{sec:arch}

\begin{figure} [t]
     \centering
%      \vspace{-1.2em}
     \includegraphics[width=0.8\columnwidth]{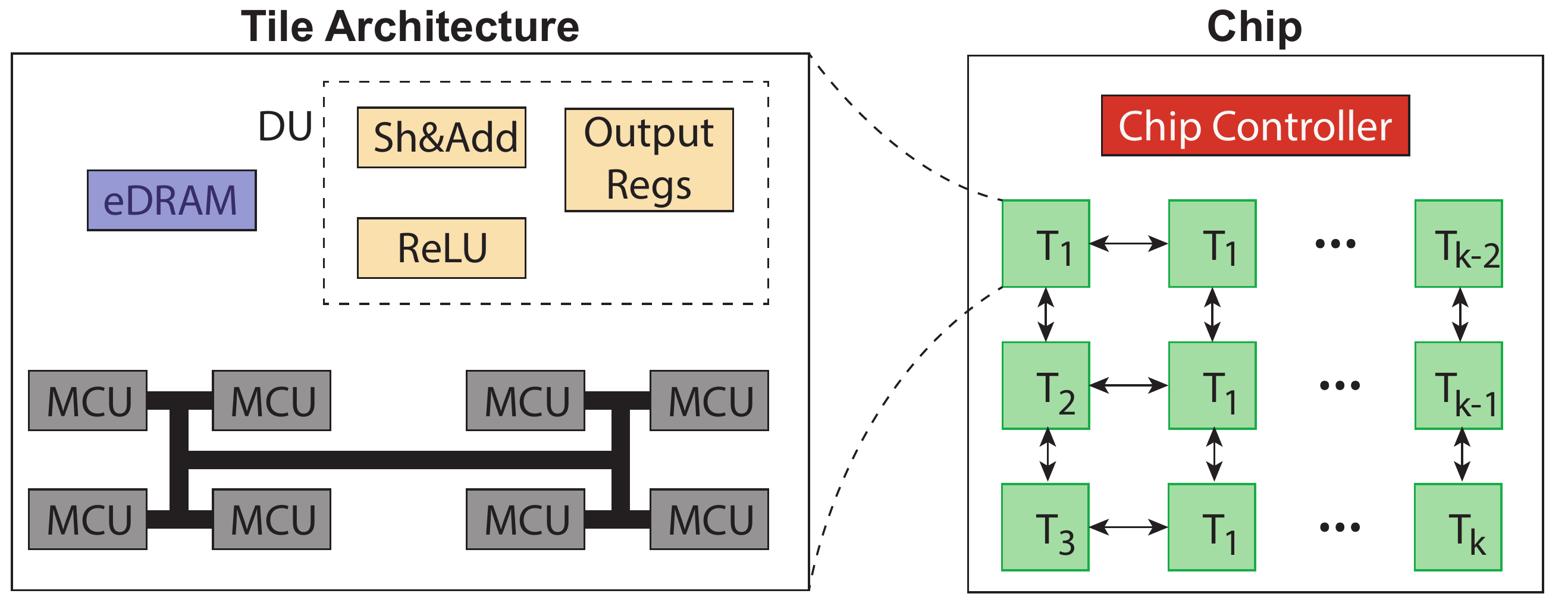}     
     \caption{Components of FORMS architecture.}
     \label{fig:circuit_architecture}
\end{figure}

\begin{figure} [b]
     \centering
%      \vspace{-1.2em}
     \includegraphics[width=1.0\columnwidth]{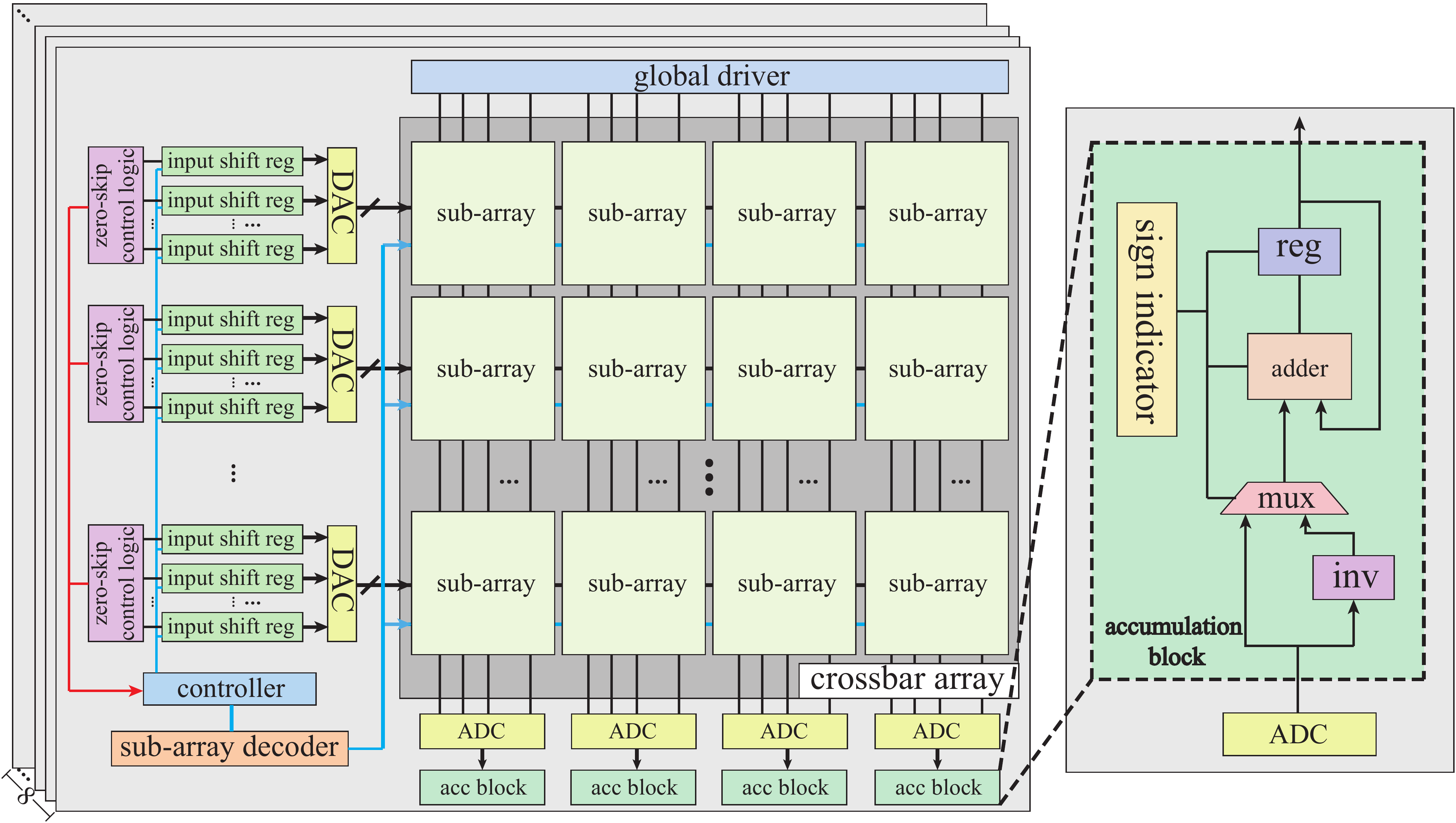}     
     \caption{FORMS MCU Architecture Design.}
     
     \label{fig:MCU_architecture}
\end{figure}

%%%%%%%%%%%%%%%%%%%%%%%%%%%%%%%%%%%%%%%%%%%%%%%%%
\begin{figure*} [t]
     \centering
%      \vspace{-1.2em}
     \includegraphics[width=0.9\textwidth]{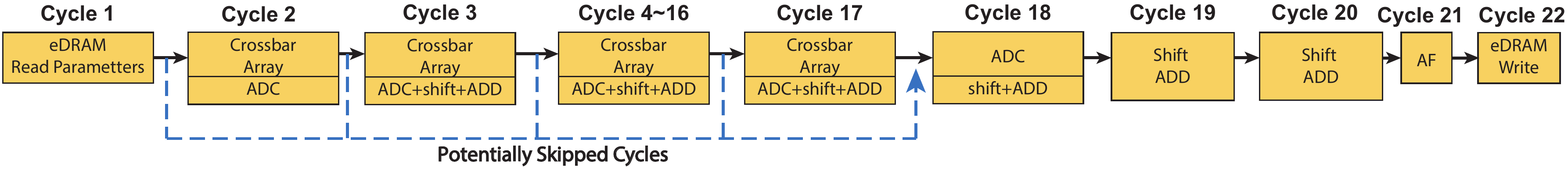}  
        %   \vspace{-0.3 in}
     \caption{{Pipeline of the FORMS Architecture.
    %  that shows flowing an operation in a layer
     }}
              \vspace{-0.2 in}
     \label{fig:pipeline}
\end{figure*}
%%%%%%%%%%%%%%%%%%%%%%%%%%%%%%%%%%%%%%%%%%

FORMS includes several architectural and circuit-level optimizations.
We performed design space exploration to find the best size of crossbar arrays, ADCs, DACs, and eDRAM storage.
The FORMS system is organized into the multiple nodes/tiles as shown in Figure~\ref{fig:circuit_architecture}. Each layer of the CNN is mapped into one or multiple tiles. Tiles are connected together in a mesh-based network while the data flow between different layers (tiles) in a pipelined manner. The chip controller orchestrates the flow of operation between different tiles. Each tile comprises multiple MAC units (MCU), eDRAMs, and digital units (DUs) that contain shift and add units, activation function, and output registers. 
Each MCU comprises eight $128 \times 128$ crossbar arrays, ADC units, and output registers. 
For fragment sizes of 16, 8, and 4, FORMS employs 5,4, and 3-bit ADCs. Unless mentioned 
explicitly, we assume the fragment size 8 in
our discussion.  
In ISAAC, the 8-bit ADCs contribute to 58\% of tile power and 31\% of tile area consumption. If with the same technology, we build a 4-bit ADC, it results in almost 4$\times$ times less area and power~\cite{murmannadc, saberi2011analysis}.
%and 1.8$\times$ times higher frequency~\cite{murmannadc, saberi2011analysis}.
By reducing the ADC resolution from 8-bit to 4-bit, the throughput will drop by a factor of 16$\times$ since each time only one fragment (i.e, 8 rows) is activated for computing dot-products using 4-bit ADCs; whereas in ISAAC, 128 rows are processed each time. 
In FORMS, we use 2-bit ReRAM cells. 
Through design space explorations, we find that 2-bit ReRAM cells delivers a better energy-efficiency than other number of bits per cell (e.g., 4-bit, 8-bit). 
ADC bits increase as we increase the ReRAM cell bits, thereby consuming more power and area.
More importantly, using more bits per ReRAM cells requires more rigorous hardware fabrication, which also introduces imprecision in analog computing and is more prone to process variation~\cite{shafiee2016isaac, xue202015}.

To increase throughput, we perform several optimizations.
For instance, ISAAC uses 8 crossbars per MCU, 1 ADC per crossbar, 12 MCUs per tile, and 168 tiles. FORMS can keep 1 ADC per crossbar, increase the number of crossbars per MCU and accordingly reduce the required MCUs and tiles. This helps us to reduce power and energy, but it cannot increase throughput. On the other side, FORMS increases the number of ADCs per crossbar to increase the throughput and yet reduce power and energy. To have an iso-area comparison with ISAAC, we increase the number of ADCs per crossbar by 4$\times$. The architecture design of the MCU is shown in Figure~\ref{fig:MCU_architecture}.
Thus, each ADC is responsible for 32 columns instead of 128 columns in ISAAC and they work in parallel, which increases the throughput by 4$\times$ compared with having 1 ADC per crossbar.

Most importantly, in ISAAC, 1 8-bit ADC should switch between 128 columns in 1 cycle (100 ns), which is infeasible in practice. FORMS architecture solves this problem by adding more but smaller ADCs with a higher sampling rate in the iso-area fashion. In addition, due to the small fragment size, the buffer size required for storing intermediate results between layers is decreased. 
%We pick the ADC sampling rate as the reference clock frequency of the design. 
The ISAAC 8-bit ADC operates at 1.2GHz and is shared by 128 columns. As a result, processing one bit of all active inputs requires ($128 \times \frac{1}{1.2GHz}=106.6ns$). Instead, FORMS employs four 4-bit ADCs (within the same area of an 8-bit ADC but 1.8$\times$ times higher frequency) to compute 128 dot-products, which results in a cycle time of $ \frac{128}{4} \times \frac{1}{2.1GHz}=$ 15 ns. As a result, FORMS improves the cycle time 
that assists to increase the throughput. Please note that we cannot scale down the ADC resolution continuously and keep improving the lower area, power, and better frequency. 
Doing so will lead to a significant system throughput drop. 
Moreover, the overhead of orchestrating the results of all small ADCs would be considerable. 

Figure~\ref{fig:pipeline} shows the proposed pipeline for FORMS. Like ISAAC, the pipeline has 22 stages (26 stages for layers that need pooling). We design a skipping logic circuit that detects the number of required cycles for shifting inputs on-the-fly. Considering that inputs are 16(8) bits, normally a total of 16(8) cycles are necessary to feed all the inputs to the crossbar array. However, FORMS does not need to wait for the whole 16(8) cycles since usually most significant bits are zeros. In most cases, we can finish feeding inputs to the crossbar early and start computation faster.
In addition, feeding zero bits wastes power and energy.
The inputs are fed into a crossbar using a shift register with the parallel-in, serial-out property. To detect the number of required cycles to shift inputs to a fragment, for the zero skipping logic, we NOR all the bits in each shift register every cycle. The results of the NOR operations per each fragment are AND-ed and used to trigger ADC to start the computation. As soon as contents of all shift registers become zero, the output of the AND is ``1", shifting the 8-bit values stops and the ADC starts computing the results of the dot-product. 
In contrast to ISAAC, the number of iterations for inputs is variable as FORMS skips some cycle of execution if all the input bits are zero, which happens frequently due to the small size of the fragment,
The results of the ADC logic is delivered to the ``Shift and ADD" units for accumulation in the output registers. They are fed into the activation function and results are stored into an eDRAM, which are inputs for the next layer. When we perform pooling, eDRAM is read in cycles 23-36 to perform max-pooling. In this case, the max of 4 values is computed and written back into eDRAM.

\section{Evaluation Results}  
\label{sec:5}
In this section, we first evaluate our proposed FORMS optimization framework in terms of model accuracy and ReRAM crossbar reduction achieved by combining crossbar-aware structured pruning, polarization, and quantization techniques. 
And we compare our results with several representative pruning works. 
Then, we analyze our FORMS accelerator architecture design in terms of area, power, throughput, and performance rate, and compare the results with the state-of-the-art accelerator designs.
At the end of this section, we analyze the impact of variation from both the software and hardware perspective.
All the results of FORMS are based on 16-bit inputs/activations and 8-bit weights, and 2-bit ReRAM cells are used.
The detailed experimental setups for the software and hardware results are illustrated in Section~\ref{sec:result_software} and Section~\ref{sec:result_hardware}.

% ------------- MNIST and CIFAR-10 -------------
\begin{table}[t]
\footnotesize
\caption{The experimental results of FORMS on multi-layer network on small to medium datasets.}
\centering
\scalebox{0.8}{
\begin{tabular}{|c|c|c|c|c|c|}
\hline
Method & \begin{tabular}[c]{@{}c@{}}Original\\ Acc.\\ (32-bit)\end{tabular} & \begin{tabular}[c]{@{}c@{}}Prune\\ Ratio\end{tabular} & \begin{tabular}[c]{@{}c@{}}Fragment\\ Size\end{tabular} & \begin{tabular}[c]{@{}c@{}}Acc.\\Drop\\ (8-bit)\end{tabular} & \begin{tabular}[c]{@{}c@{}}Crossbar\\ Reduction\end{tabular} \\ \hline
\multicolumn{6}{|c|}{MNIST} \\ \hline
\begin{tabular}[c]{@{}c@{}}GroupScissor~\cite{wang2017group} \\ LeNet5\end{tabular} & 99.15\% & 4.2$\times$ & - &
\begin{tabular}[c]{@{}c@{}}0.01\%\\ (32-bit)\end{tabular} & 4.2$\times$ \\ \hline
\multirow{3}{*}{\textbf{\begin{tabular}[c]{@{}c@{}}our\\ LeNet5\end{tabular}}} & \multirow{3}{*}{\textbf{99.17\%}} & \multirow{3}{*}{\textbf{23.18$\times$}} & \textbf{4} & \textbf{-0.02\%} & \multirow{3}{*}{\textbf{185.44$\times$}} \\ \cline{4-5}
 &  &  & \textbf{8} & \textbf{-0.01\%} &  \\ \cline{4-5}
 &  &  & \textbf{16} & \textbf{0.14\%} &  \\ \hline
\multicolumn{6}{|c|}{CIFAR-10} \\ \hline
\begin{tabular}[c]{@{}c@{}}IterativePrune~\cite{han2015learning} \\ VGG16\end{tabular} & 92.50\% & 2$\times$ & - & \begin{tabular}[c]{@{}c@{}}0.3\%\\ (32-bit)\end{tabular} & 2$\times$ \\ \hline
%-------- tiny but acc vgg16 ---------
\begin{tabular}[c]{@{}c@{}}TinyButAcc~\cite{tiny_but_accurate} \\ VGG16\end{tabular} & 93.70\% & 44.67$\times$ & - & \begin{tabular}[c]{@{}c@{}}0.66\%\end{tabular} & 164.8$\times$ \\ \hline
%-------- our vgg16 ---------
\multirow{3}{*}{\textbf{\begin{tabular}[c]{@{}c@{}}our\\ VGG16\end{tabular}}} & \multirow{3}{*}{\textbf{93.70\%}} & \multirow{3}{*}{\textbf{41.2$\times$}} & \textbf{4} & \textbf{0.61\%} & \multirow{3}{*}{\textbf{329.6$\times$}} \\ \cline{4-5}
 &  &  & \textbf{8} & \textbf{0.64\%} &  \\ \cline{4-5}
 &  &  & \textbf{16} & \textbf{0.77\%} &  \\ \hline
 %-------- AMC resnet18 ---------
\begin{tabular}[c]{@{}c@{}}AMC~\cite{he2018amc} \\ResNet18\end{tabular} & 90.5\% & 2$\times$ & - & \begin{tabular}[c]{@{}c@{}}0.3\%\\ (32-bit)\end{tabular} & 2$\times$ \\ \hline
%-------- Tiny resnet18 ---------
\begin{tabular}[c]{@{}c@{}}TinyButAcc~\cite{tiny_but_accurate} \\ ResNet18\end{tabular} & 94.14\% & 52.07$\times$ & - & \begin{tabular}[c]{@{}c@{}}0.35\%\end{tabular} & 203.4$\times$ \\ \hline
%-------- our resnet18 ---------
\multirow{3}{*}{\textbf{\begin{tabular}[c]{@{}c@{}}our\\ ResNet18\end{tabular}}} & \multirow{3}{*}{\textbf{94.14\%}} & \multirow{3}{*}{\textbf{50.85$\times$}} & \textbf{4} & \textbf{0.35\%} & \multirow{3}{*}{\textbf{406.8$\times$}} \\ \cline{4-5}
 &  &  & \textbf{8} & \textbf{0.47\%} &  \\ \cline{4-5}
 &  &  & \textbf{16} & \textbf{0.92\%} &  \\ \hline
\end{tabular}
}   % scalebox
\label{table:MNIST_CIFAR}
\end{table}

% ------------- CIFAR-100 and ImageNet -------------
\begin{table}[t]
% \footnotesize
\caption{The experimental results of FORMS on multi-layer network on medium to large datasets. Top-5 accuracy is used for ImageNet dataset.}
\centering
\scalebox{0.8}{
\begin{tabular}{|c|c|c|c|c|c|}
\hline
Method & \begin{tabular}[c]{@{}c@{}}Original\\ Acc.\\ (32-bit)\end{tabular} & \begin{tabular}[c]{@{}c@{}}Prune\\ Ratio\end{tabular} & \begin{tabular}[c]{@{}c@{}}Fragment\\ Size\end{tabular} & \begin{tabular}[c]{@{}c@{}}Acc.\\ Drop\\ (8-bit)\end{tabular} & \begin{tabular}[c]{@{}c@{}}Crossbar\\ Reduction\end{tabular} \\ \hline
\multicolumn{6}{|c|}{CIFAR-100} \\ \hline
% ------------- FPGM resnet20 cifar100 -------------
\begin{tabular}[c]{@{}c@{}}FPGM~\cite{FPGM} \\ ResNet20\end{tabular} & 67.62\% & 1.73$\times$ & - & \begin{tabular}[c]{@{}c@{}}0.76\%\\ (32-bit)\end{tabular} & 1.73$\times$ \\ \hline
% ------------- our resnet18 cifar100 -------------
\multirow{3}{*}{\textbf{\begin{tabular}[c]{@{}c@{}}our\\ ResNet18\end{tabular}}} & \multirow{3}{*}{\textbf{76.37\%}} & \multirow{3}{*}{\textbf{6.65$\times$}} & \textbf{4} & \textbf{-0.06\%} & \multirow{3}{*}{\textbf{53.2$\times$}} \\ \cline{4-5}
 &  &  & \textbf{8} & \textbf{-0.03\%} &  \\ \cline{4-5}
 &  &  & \textbf{16} & \textbf{0.17\%} &  \\ \hline
% ------------- FPGM resnet56 cifar100 -------------
\begin{tabular}[c]{@{}c@{}}FPGM~\cite{FPGM} \\ ResNet56\end{tabular} & 71.41\% & 2.11$\times$ & - & \begin{tabular}[c]{@{}c@{}}1.75\%\\ (32-bit)\end{tabular} & 2.11$\times$ \\ \hline
% ------------- our resnet50 cifar100 -------------
\multirow{3}{*}{\textbf{\begin{tabular}[c]{@{}c@{}}our\\ ResNet50\end{tabular}}} & \multirow{3}{*}{\textbf{77.35\%}} & \multirow{3}{*}{\textbf{9.18$\times$}} & \textbf{4} & \textbf{0.10\%} & \multirow{3}{*}{\textbf{73.44$\times$}} \\ \cline{4-5}
 &  &  & \textbf{8} & \textbf{0.31\%} &  \\ \cline{4-5}
 &  &  & \textbf{16} & \textbf{0.61\%} &  \\ \hline
% ------------- Network slime cifar100 -------------
\begin{tabular}[c]{@{}c@{}}Network Slim~\cite{liu2017learning}\\ VGG16\end{tabular} & 73.26\% & 2$\times$ & - & \begin{tabular}[c]{@{}c@{}}-0.22\%\\ (32-bit)\end{tabular} & 2$\times$ \\ \hline
% ------------- our vgg16 cifar100 -------------
\multirow{3}{*}{\textbf{\begin{tabular}[c]{@{}c@{}}our\\ VGG16\end{tabular}}} & \multirow{3}{*}{\textbf{73.32\%}} & \multirow{3}{*}{\textbf{8.15$\times$}} & \textbf{4} & \textbf{-0.01\%} & \multirow{3}{*}{\textbf{65.20$\times$}} \\ \cline{4-5}
 &  &  & \textbf{8} & \textbf{0.10\%} &  \\ \cline{4-5}
 &  &  & \textbf{16} & \textbf{0.37\%} &  \\ \hline
\multicolumn{6}{|c|}{ImageNet} \\ \hline
% ------------- DCP Imagenet -------------
\begin{tabular}[c]{@{}c@{}}DCP~\cite{zhuang2018discrimination} \\ ResNet18\end{tabular} & 88.98\% & 1.42$\times$ & - & \begin{tabular}[c]{@{}c@{}}0.12\%\\ (32-bit)\end{tabular} & 1.42$\times$ \\ \hline
% ------------- Tiny Imagenet -------------
\begin{tabular}[c]{@{}c@{}}TinyButAcc~\cite{tiny_but_accurate} \\ ResNet18\end{tabular} & 89.07\% & 3.33$\times$ & - & \begin{tabular}[c]{@{}c@{}}0.6\%\end{tabular} & 12.4$\times$ \\ \hline
% ------------- our imagenet -------------
\multirow{6}{*}{\textbf{\begin{tabular}[c]{@{}c@{}}our\\ ResNet18\end{tabular}}} & \multirow{6}{*}{\textbf{89.08\%}} & \multirow{3}{*}{\textbf{1.67$\times$}} & \textbf{4} & \textbf{0.03\%} & \multirow{3}{*}{\textbf{13.36$\times$}} \\ \cline{4-5}
 &  &  & \textbf{8} & \textbf{0.27\%} &  \\ \cline{4-5}
 &  &  & \textbf{16} & \textbf{1.19\%} & \\ \cline{3-6} 
 &  & \multirow{3}{*}{\textbf{2.0$\times$}} & \textbf{4} & \textbf{0.34\%} & \multirow{3}{*}{\textbf{16.0$\times$}} \\ \cline{4-5}
 &  &  & \textbf{8} & \textbf{0.62\%} &  \\ \cline{4-5}
 &  &  & \textbf{16} & \textbf{1.73\%} & \\ \hline
 % ------------- our imagenet ResNet-50 -------------
\multirow{6}{*}{\textbf{\begin{tabular}[c]{@{}c@{}}our\\ ResNet50\end{tabular}}} & \multirow{6}{*}{\textbf{92.34\%}} & \multirow{3}{*}{\textbf{2.15$\times$}} & \textbf{4} & \textbf{0.13\%} & \multirow{3}{*}{\textbf{17.2$\times$}} \\ \cline{4-5}
 &  &  & \textbf{8} & \textbf{0.34\%} &  \\ \cline{4-5}
 &  &  & \textbf{16} & \textbf{1.17\%} & \\ \cline{3-6} 
 &  & \multirow{3}{*}{\textbf{3.67$\times$}} & \textbf{4} & \textbf{0.37\%} & \multirow{3}{*}{\textbf{29.36$\times$}} \\ \cline{4-5}
 &  &  & \textbf{8} & \textbf{0.70\%} &  \\ \cline{4-5}
 &  &  & \textbf{16} & \textbf{1.62\%} & \\ \hline
\end{tabular}
}   % scalebox
\label{table:CIFAR100_IMAGENET}
\vspace{3mm}
\end{table}

% ##################################################
\begin{table}[t]
% \normalsize
\Huge
% \vspace{-3mm}
\caption{FORMS MCU hardware specification and comparison with ISAAC.}\label{table:PAresult}
\centering
\scalebox{0.48}{
% \begin{tabular}{p{1.8cm}p{1.4cm}p{1.4cm}p{1.2cm}p{1.2cm}p{1.4cm}p{1.2cm}p{1.2cm}}
\resizebox{\textwidth}{!}{
\begin{tabular}{c c c c c c c c}
\hline
\hline
%------------- 1st row -----------------------
\multicolumn{2}{c|}{} & 
\multicolumn{3}{|c|}{\textbf{FORMS (Fragment size 8)}} & \multicolumn{3}{|c}{\textbf{ISAAC~\cite{shafiee2016isaac}}} \\
\hline
%------------- 2nd row -----------------------
%\multicolumn{2}{c|}{} & \multicolumn{1}{|c}{} & Power($mW$) & \multicolumn{1}{c|}{Area($mm^2$)}  & \multicolumn{1}{|c}{} & Power($mW$) & Area($mm^2$)\\ \hline
%DU unit  & \multicolumn{1}{c|}{}  & \multicolumn{1}{|c}{} & 53.05 & \multicolumn{1}{c|}{0.25}  & \multicolumn{1}{|c}{} & 40.85 & 0.213\\ \hline
% MCU unit  & \multicolumn{1}{c|}{}  & \multicolumn{1}{|c}{} & 1111 & \multicolumn{1}{c|}{1111} & \multicolumn{1}{|c}{} & 1111 & 1111\\ \hline
 \hline
%---------------- Components --------------------------------
Component & \multicolumn{1}{c|}{Parameter} & \multicolumn{1}{|c}{Spec} & Power($mW$) & \multicolumn{1}{c|}{Area($mm^2$)} & \multicolumn{1}{|c}{Spec} & Power($mW$) & Area($mm^2$)\\ \hline
% ------------------- ADC ------------------------
 \multirow{3}{*}{\makecell{ADC}} &  \multicolumn{1}{c|}{\multirow{3}{*}{\makecell{resolution \\ frequency \\ number}}} &
 \multicolumn{1}{|c}{\multirow{3}{*}{\makecell{ 4-bit \\ 2.1GHz \\ 32}}} &
 \multirow{3}{*}{\makecell{ 15.2 }} &
 \multicolumn{1}{c|}{\multirow{3}{*}{\makecell{ 0.0091 }}} &
 \multicolumn{1}{|c}{\multirow{3}{*}{\makecell{ 8-bit \\ 1.2GHz \\ 8}}} &
 \multirow{3}{*}{\makecell{ 16 }} &
 \multirow{3}{*}{\makecell{ 0.0096 }} \\ 
 & \multicolumn{1}{c|}{} & \multicolumn{1}{|c}{} & & \multicolumn{1}{c|}{} & \multicolumn{1}{|c}{} &&\\
 & \multicolumn{1}{c|}{} & \multicolumn{1}{|c}{} & & \multicolumn{1}{c|}{} & \multicolumn{1}{|c}{} &&\\
\hline
% ------------------- DAC ------------------------
 \multirow{2}{*}{\makecell{DAC}} &  \multicolumn{1}{c|}{\multirow{2}{*}{\makecell{resolution \\ number}}} &
 \multicolumn{1}{|c}{\multirow{2}{*}{\makecell{1-bit \\ 8$\times$128}}} &
 \multirow{2}{*}{\makecell{4}} &
 \multicolumn{1}{c|}{\multirow{2}{*}{\makecell{0.00017}}} &
 \multicolumn{1}{|c}{\multirow{2}{*}{\makecell{1-bit \\ 8$\times$128}}} &
 \multirow{2}{*}{\makecell{4}} &
 \multirow{2}{*}{\makecell{0.00017}} \\
 & \multicolumn{1}{c|}{} & \multicolumn{1}{|c}{} & & \multicolumn{1}{c|}{} & \multicolumn{1}{|c}{} &&\\
 \hline
 % ------------------- S&H ------------------------
 S\&H & \multicolumn{1}{c|}{number} & \multicolumn{1}{|c}{8$\times$128} & 0.0055 & \multicolumn{1}{c|}{0.000023} & 
 \multicolumn{1}{|c}{8$\times$128} & 0.01 & 0.00004 \\
 \hline
 
 % ------------------- crossbar array ------------------------
 \multirow{3}{*}{\makecell{crossbar array}} & 
 \multicolumn{1}{c|}{\multirow{3}{*}{\makecell{number \\ size \\ bit/cell}}} &
 \multicolumn{1}{|c}{\multirow{3}{*}{\makecell{ 8 \\ 128$\times$128 \\ 2}}} &
 \multirow{3}{*}{\makecell{ 2.44}} &
 \multicolumn{1}{c|}{\multirow{3}{*}{\makecell{ 0.00024}}} &
 \multicolumn{1}{|c}{\multirow{3}{*}{\makecell{ 8 \\ 128$\times$128 \\ 2}}} &
 \multirow{3}{*}{\makecell{ 2.43}} &
 \multirow{3}{*}{\makecell{ 0.00023}} \\
 & \multicolumn{1}{c|}{} & \multicolumn{1}{|c}{} & & \multicolumn{1}{c|}{} & \multicolumn{1}{|c}{} &&\\
 & \multicolumn{1}{c|}{} & \multicolumn{1}{|c}{} & & \multicolumn{1}{c|}{} & \multicolumn{1}{|c}{} &&\\
 \hline
  % ------------------- S+A ------------------------
 S+A & \multicolumn{1}{c|}{number} & \multicolumn{1}{|c}{4} & 0.2 & \multicolumn{1}{c|}{0.000024} & 
 \multicolumn{1}{|c}{4} & 0.2 & 0.000024 \\
 \hline
% ------------------- Skipping logic ------------------------
 Skipping Logic & \multicolumn{1}{c|}{} & \multicolumn{1}{|c}{} & 0.01 & \multicolumn{1}{c|}{0.0000001} & 
 \multicolumn{1}{|c}{} & -- & -- \\
 \hline
 Sign Indicator & \multicolumn{1}{c|}{} & \multicolumn{1}{|c}{} & 0.012 & \multicolumn{1}{c|}{0.0000031} & 
 \multicolumn{1}{|c}{} & -- & -- \\
 \hline

 \hline

\end{tabular}
}
}
 %\vspace{-2mm}
\end{table}

% ##################################################
\begin{table*}[t]
\centering
\footnotesize
% \vspace{-3mm}
\caption{Comparison of FORMS hardware characteristics with ISAAC and DaDianNao}\label{table:PAresultsummary}
\resizebox{\textwidth}{!}{
\begin{tabular}{c c c c c c c c c c c}
\hline
  \hline
  \multicolumn{11}{c}{\textbf{Total}} \\
  %------------- 1st row -----------------------
  \hline
\multicolumn{1}{c|}{} & 
\multicolumn{3}{|c|}{\textbf{FORMS (Fragment size 8)}} & \multicolumn{3}{|c|}{\textbf{ISAAC~\cite{shafiee2016isaac}}} &
\multicolumn{1}{|c|}{} & 
\multicolumn{3}{|c}{\textbf{DaDianNao~\cite{chen2014dadiannao}} }\\
  \hline
  \multicolumn{1}{c|}{} & \multicolumn{1}{|c}{Spec} & Power($mW$) & \multicolumn{1}{c|}{Area($mm^2$)} & \multicolumn{1}{|c}{Spec} & Power($mW$) & \multicolumn{1}{c|}{Area($mm^2$)} & \multicolumn{1}{|c|}{} &  \multicolumn{1}{|c}{Spec} & Power($mW$) & Area($mm^2$) \\ \hline
  
 \multicolumn{1}{c|}{Dig unit}  & \multicolumn{1}{|c}{} & 53.05 & \multicolumn{1}{c|}{0.25}  & \multicolumn{1}{c}{} &   40.85 & \multicolumn{1}{c|}{0.213} & \multicolumn{1}{c|}{} &
 \\ \hline
  \multicolumn{1}{c|}{12 MCUs per tile}  & \multicolumn{1}{|c}{} & 280.05 &  \multicolumn{1}{c|}{0.152} &  \multicolumn{1}{|c}{} & 288.96 & \multicolumn{1}{c|}{0.1580} & \multicolumn{1}{|c|}{NFU} &  \multicolumn{1}{|c}{16} & 4886 & 16.09\\ \hline
  \multicolumn{1}{c|}{1 Tile (12 MCUs + Dig unit)} &  \multicolumn{1}{|c}{} & 333.1 & \multicolumn{1}{c|}{0.39} & \multicolumn{1}{|c}{} & 329.81 & \multicolumn{1}{c|}{0.370} & \multicolumn{1}{|c|}{EDRAM} &  \multicolumn{1}{|c}{4/tile 36 MB} & 4760 & 33.12 \\ \hline
  \multicolumn{1}{c|}{168 Tiles} &  \multicolumn{1}{|c}{} & 55960.8 & \multicolumn{1}{c|}{66.27} & \multicolumn{1}{|c}{} & 55408.08 & \multicolumn{1}{c|}{62.21} & \multicolumn{1}{|c|}{Global Bus} &  \multicolumn{1}{|c}{128 bits} & 12.8 & 15.66\\ \hline
  \multicolumn{1}{c|}{HyperTransport freq} & \multicolumn{1}{|c}{4/1.6GHz} & 10400 & \multicolumn{1}{c|}{22.88} & \multicolumn{1}{|c}{4/1.6GHz} & 10400 & \multicolumn{1}{c|}{22.88} & \multicolumn{1}{|c|}{HyperTransport freq} &  \multicolumn{1}{|c}{4/1.6GHz} & 10400 & 22.88\\ 
  \multicolumn{1}{c|}{HyperTransport bw} & \multicolumn{1}{|c}{6.4GB/s} &  & \multicolumn{1}{c|}{} & \multicolumn{1}{|c}{6.4GB/s} & & \multicolumn{1}{c|}{} & \multicolumn{1}{|c|}{HyperTransport bw} &  \multicolumn{1}{|c}{6.4 GB/s} &  & \\ \hline
 \multicolumn{1}{c|}{ Chip Total} & \multicolumn{1}{|c}{} & 66360.8 & \multicolumn{1}{c|}{89.15} & \multicolumn{1}{|c}{} & 65808.08 & \multicolumn{1}{c|}{85.09} & \multicolumn{1}{|c|}{Chip Total} &  \multicolumn{1}{|c}{} & 19856 & 86.2\\ \hline
  \hline

\end{tabular}
}
%  \vspace{-2mm}
\end{table*}

\subsection{Comparing Model Compression Methods}
\label{sec:result_software}

We evaluate our method using several representative benchmark networks, including LeNet-5, ResNet-18/50, and VGG-16, and using MNIST, CIFAR-10/100, and ImageNet datasets.
The crossbar reduction results are from our crossbar-aware structured pruned, polarized, and quantized models comparing to the original baseline model using a splitting mapping scheme~\cite{sayyaparaju2017circuit}. 
All models are trained on an 8$\times$ NVIDIA Quadro RTX 6000 GPU server by PyTorch API.

\emph{\textbf{MNIST \& CIFAR-10.}} 
Table~\ref{table:MNIST_CIFAR} shows, on MNIST dataset using LeNet5, FORMS achieves 185.44$\times$ crossbar reduction, where 23.18$\times$ reduction comes from crossbar-aware structured pruning, 4$\times$ reduction comes from quantization, and 2$\times$ reduction from polarization by eliminating half of the crossbars to represent positive/negative weights. When fragment size is 4 or 8, there is no accuracy degradation (in fact, the accuracy is higher than the original model since the pruning may reduce the overfitting), where there is a minor accuracy loss for the fragment size of 16. FORMS achieves a much higher crossbar reduction ratio than \textsc{GroupScissor}~\cite{wang2017group}.
On CIFAR-10 dataset, we compare our results with several reference works. \textsc{TinyButAcc}~\cite{tiny_but_accurate} is a state-of-the-art ReRAM-based pruning work using ADMM. 
By taking the crossbar size into account, we achieve a similar crossbar reduction ratio from the pruning part as \textsc{TinyButAcc}, with a lower pruning ratio.
This helps FORMS avoid unnecessary accuracy drop. 
By combining the fragment polarization and quantization, our overall crossbar reduction ratio is 2$\times$ higher than \textsc{TinyButAcc} with similar accuracy.

\emph{\textbf{CIFAR-100 \& ImageNet.}} 
Compared to CIFAR-10 dataset, the classification task on CIFAR-100 and ImageNet dataset is more complicated. 
Especially for ImageNet, there are fewer redundant weights. 
Thus, it is common to have a much smaller pruning ratio on CIFAR-100 and Imagenet dataset.
As shown in Table~\ref{table:CIFAR100_IMAGENET}, on CIFAR-100 dataset, without accuracy loss or with minor accuracy loss, we achieve 6.65$\times$(53.2$\times$), 9.18$\times$(73.44$\times$) and 8.15$\times$(65.20$\times$) weight pruning ratio (crossbar reduction) on ResNet-18, ResNet-50 and VGG-16, respectively.

On ImageNet, we compare our results with references on ResNet-18. Since model accuracy is more sensitive to pruning ratio on ImageNet dataset, to achieve a higher overall crossbar reduction ratio and maintain model accuracy, we use a less aggressive pruning strategy.
Compared to \textsc{TinyButAcc}, 
we achieve higher overall crossbar reduction (13.36$\times$ and 16$\times$) with higher or similar accuracy when using fragment size of 4 or 8. 
In summary, compared to state-of-the-art works, FORMS achieves high crossbar reduction with no or little accuracy loss.

%%%%%%%%%%%%%%%%%%%%%%%%%%%%%%%%%%%%%%%%%%%%%%
\subsection{Area and Power}
\label{sec:result_hardware}
We develop an in-house simulator to model access latency, energy, and area of all buffers, on-chip interconnects and crossbar arrays as well as the performance of FORMS. The back-end of the tool utilizes CACTI 7.0 ~\cite{balasubramonian2017cacti}, NVSIM, and NVSIM-CAM ~\cite{li2016nvsim, dong2012nvsim} to provide a unified platform that can support both volatile and
non-volatile memories with multi-banking properties.
It can also model process variations of the size and the threshold voltage of transistors. We conservatively choose a 10\% process variation for evaluations. We use the VTEAM ReRAM model~\cite{kvatinsky2015vteam}. The zero skipping logic is modeled in Verilog HDL and synthesized using Synopsys Design Compiler at 45nm technology and scaled down to 32nm. 
We follow the methodology of ISAAC paper to model max-pooling, shift-and-add, ADC, DACs, and activation functions. The energy and area of the max-pool and shift-and-add are adapted from ISAAC~\cite{shafiee2016isaac} while for ReLU activation function circuits we use the information provided in PRIME~\cite{PRIME}. 
A charge pump circuit~\cite{palumbo2010charge} is used to provide higher voltage as ReRAM cells require a write voltage more than the supply Vdd.  The off-chip links are the same HyperTransport serial link used by ISAAC~\cite{shafiee2016isaac} and DaDianNao~\cite{chen2014dadiannao}.
Also, the power and area of 1-bit DAC (a simple inverter) are adopted from~\cite{saberi2011analysis}. 
The ADC energy and area are taken from~\cite{murmannadc}. To get power and area of the same style ADC used in ISAAC, but with 4-bit resolution, we scale down the power and area of the memory, clock, and \textit{vref} buffer linearly, and the capacitive DAC exponentially~\cite{saberi2011analysis}. The same scaling has been used in other works like ISAAC.
We utilize a 4-bit ADC with 2.1 GSps~\cite{chan201716}. We choose this methodology to model peripheral circuitry and make a fair comparison at 32 nm with state-of-the-art works.
We also compare the results with the fully digital DaDianNao~\cite{chen2014dadiannao}, after scaling from 28nm to 32nm.

% \vspace{2ex}
\begin{table}[t]
\footnotesize
% 	\vspace{-1ex}
	\caption{Comparison the effective peak nominal throughput per area unit and power unit of different architectures normalized to ISAAC.}
	\label{table:GOPS}
% 	\vspace{-2ex}
	\centerline{
		\scalebox{1.1}{
	{\scriptsize\setlength\tabcolsep{9pt}
		\begin{tabular}{|c|c|c|}
			\hline
			\textbf{Architecture}  &\textbf{$\frac {GOPs}{s \times mm^{2}}$}  & \textbf{$\frac {GOPs}{W}$} \\
			\hline
			\hline
			ISAAC & 1 & 1   \\
			DaDianNao & 0.13 & 0.45   \\ 
			PUMA & 0.70 & 0.79   \\
			TPU &0.08 & 0.48 \\
			WAX\cite{gudaparthi2019wire} & 0.33 & 2.3 \\
			SIMBA\cite{shao2019simba} & 0.34 & 0.08-2.5 \\
			FORMS (polarization only, 8) & 0.54 & 0.61 \\
			FORMS (polarization only, 16) & 0.77 & 0.84 \\
			Pruned/Quantized-ISAAC & 26.4 & 26.61   \\
			Pruned/Quantized-PUMA & 18.67 & 21.07   \\
			FORMS (full optimization, 8) & 36.02 & 27.73 \\
			FORMS (full optimization, 16) & 39.48 & 51.26 \\
			\hline
		\end{tabular}}
	}
}
% \vspace{-1ex}
\end{table}

To make a fair ISO-area comparison on throughput and frame processing rate with ISAAC, 
we build FORMS to have similar power and area to ISAAC.
We compare the power and area of the proposed building blocks in Table~\ref{table:PAresult}. Table~\ref{table:PAresultsummary} shows the area and power of the main components of FORMS, ISAAC and DaDianNao (digital design) without incorporating the pruning, and quantization, while the fragment is set to 8. 
Specifically, FORMS needs sign indicator arrays, and a few interconnects, which connect ADCs to fragments.  
It is noteworthy to mention that polarization help to accommodate more weights in the same crossbar than existing techniques. 
In addition, since FORMS performs more computations than ISAAC, it needs higher bandwidth (512-bit versus 256-bit in ISAAC) and larger eDRAM to store the results(128KB vs. 64KB in ISAAC). 
However, since an ADC in FORMS generates 16 levels compared with 256 levels generated in ISAAC (due to using small fragment sizes), the required sample\&hold circuit is smaller and faster.
Overall, the sample\&hold circuit in FORMS is almost $2\times$ smaller than ISAAC.
In addition, Zero-skipping logic saves dynamic power consumption by feeding fewer input bits (useless 0s) to the crossbar.
In summary, we build FORMS to have almost the same power and area as ISAAC. The difference is negligible (0.08\% more power and 4.5\% more area).
Like ISAAC and PUMA, in return for consuming more area and power compared with DaDianNao, the throughput of FORMS is increased significantly.

\subsection{Throughput}
\label{sec:Throughputsec}
Table~\ref{table:GOPS} compares the peak nominal area-, power-efficiency of FORMS that uses the fragment size of 8 and 16 with state-of-the-art DNN accelerators in terms of the number of operations performed per second per $mm^{2}$ and the number of operations performed per watt (i.e. $\frac{GOPs}{s \times mm^{2}}$ and $\frac{GOPs}{w}$). The results are normalized to the ISAAC in terms of  $\frac{GOPs}{s \times mm^{2}}$ and $\frac{GOPs}{w}$, respectively. As results demonstrate, by taking advantage of the proposed pruning and quantization methods, the throughput of the ISAAC and PUMA are increased considerably. This potential increase can be achieved if interconnects can provide enough bandwidth for the processing crossbars.     
By only considering the polarization, the effective peak throughput of FORMS with the fragment size of 8 is 4.15$\times$(1.36$\times$) and 6.75$\times$(1.27$\times$) over DaDianNao and TPU, respectively. By increasing the fragment size, the throughput is increased. For example, when we increase the fragment size to 16, the throughput of FORMS with the polarization only is increased by 42\%(29\%). FORMS can outperform existing architectures when we apply all of the proposed methods including pruning, quantization, polarization, and zero-skipping logic. In terms of $\frac{GOPs}{s \times mm^{2}}$, FORMS with the fragment size of 16, outperforms the original non-pruned ISAAC and Pruned/Quantized ISAAC by 39$\times$ and 1.5$\times$, respectively. These results in terms of $\frac{GOPs}{w}$ are 51$\times$ and 1.9$\times$. Overall, the peak throughput of the FORMS is higher than the existing designs.
Note that WAX and SIMBA trade-off throughput for power efficiency by reducing voltage and frequency. For instance, in SIMBA, the voltage is 0.48V and frequency is 0.52GHz, whereas, in WAX, the frequency is 0.2GHz while the voltage is not mentioned.

\subsection{Frame Processing Rate}
Reaching the higher throughput per area unit and power unit is not the only goal of an acceleration design~\cite{jouppi2017datacenter,shao2019simba}. For instance, the throughput per area efficiency of commercial TPU ~\cite{jouppi2017datacenter, ankit2019puma} is only 41 $\frac{GOPs}{s \times mm^{2}}$.    
Moreover, many DNN-based applications take advantage of a higher frame per second (fps) speed up or end to end throughput. For examples, object detection~\cite{redmon2016you}, GoTurn~\cite{held2016learning}, self-driving cars\cite{lin2018architectural}, virtual and augmented reality applications~\cite{lincoln2016motion} tend to have higher fps processing rate rather than higher throughput per area or power efficiency. This makes the high fps another crucial property for DNN accelerators.

\begin{figure} [t!]
     \centering
      \vspace{-1.2em}
     \includegraphics[width=1\columnwidth]{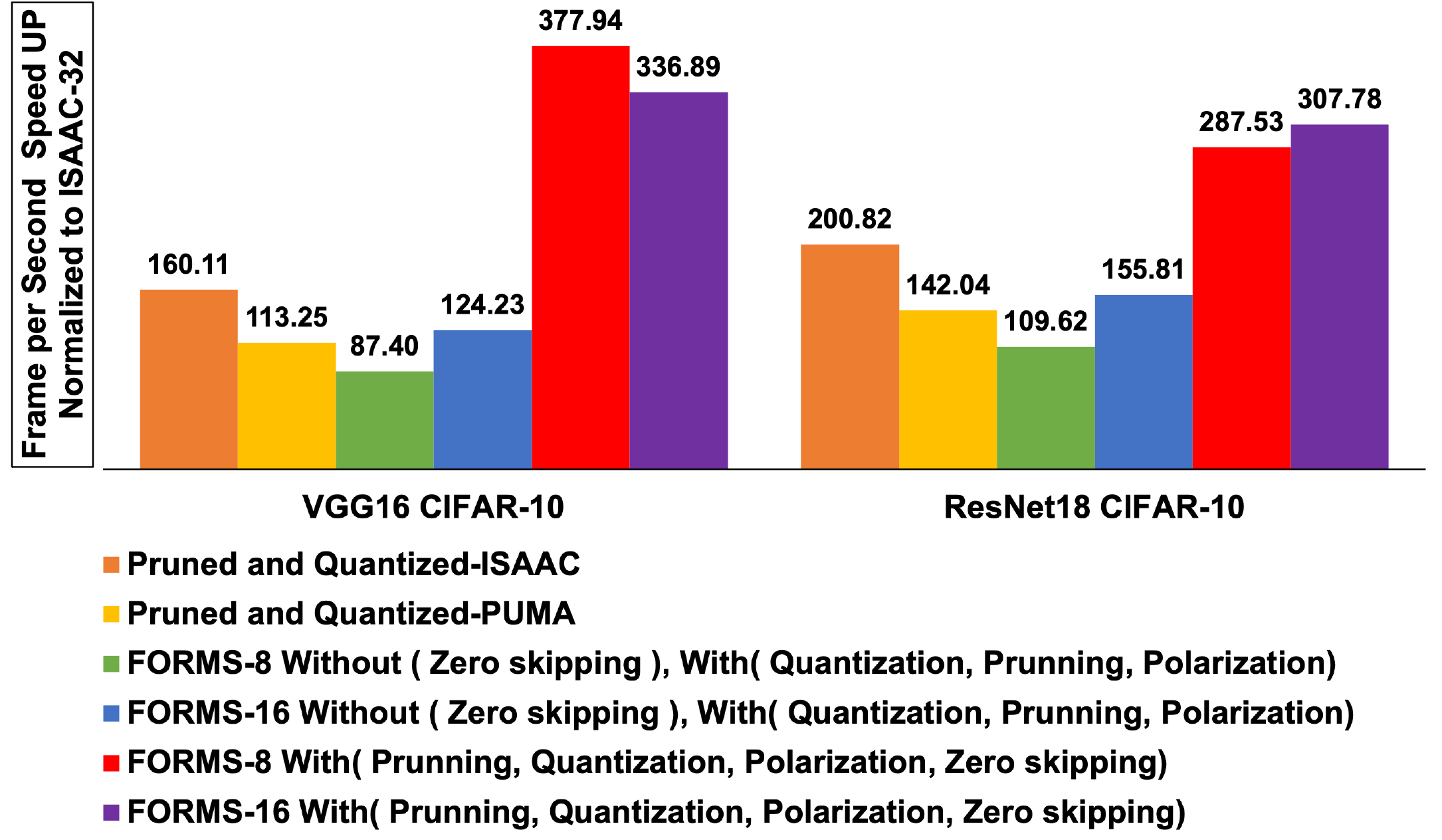}    
     \caption{Speed up results in terms of frame per second on CIFAR-10 when various techniques proposed in FORMS are applied.}
     \label{fig:SpeedupFPS1}
 %          \vspace{-1.2em}

 \end{figure}

\begin{figure} [t!]
     \centering
%      \vspace{-1.2em}
     \includegraphics[width=\columnwidth]{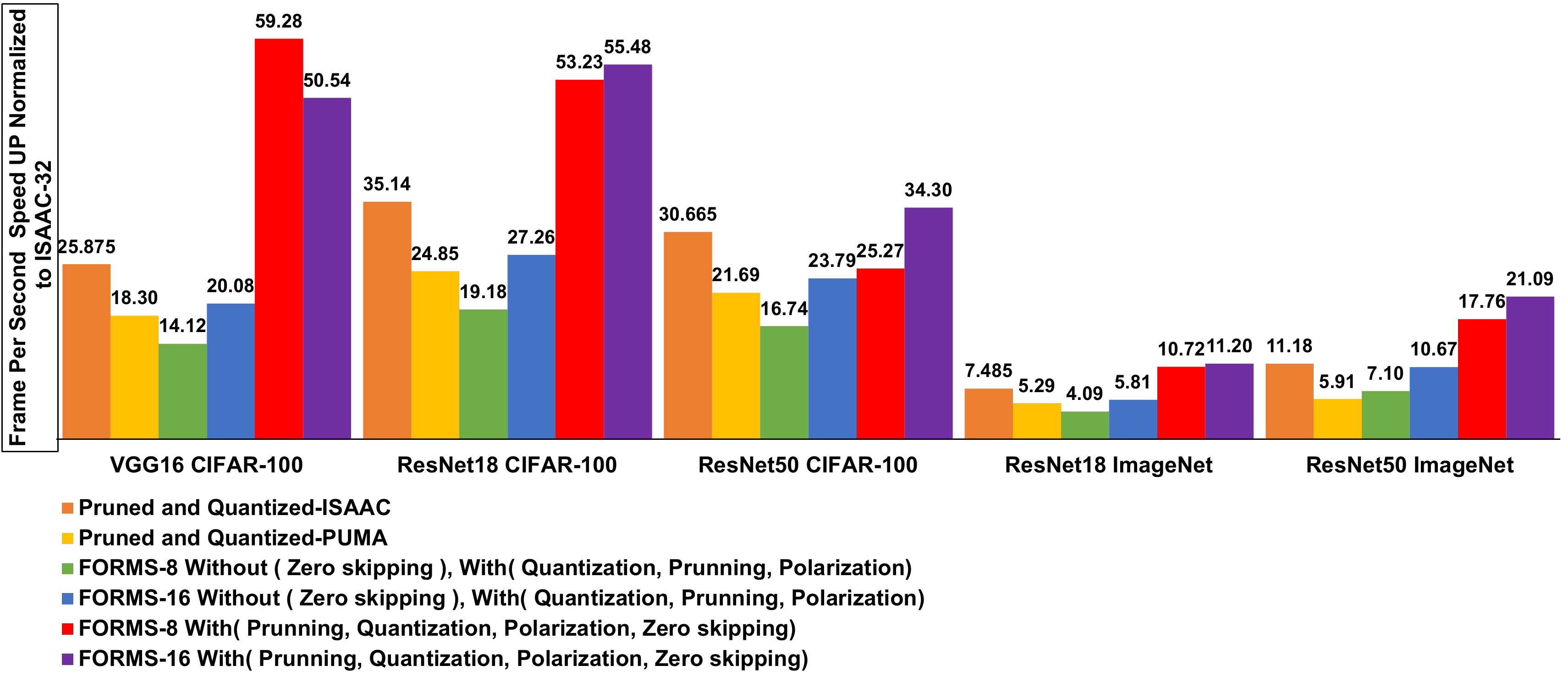}    
     \caption{Speed up results in terms of frame per second on CIFAR-100 and ImageNet when various techniques proposed in FORMS are applied.}
     \label{fig:SpeedupFPS2}
    % \vspace{1.5em}

\end{figure}

We compare the frame per second speed up of different architectures when we apply assorted proposed techniques. 
Figure~\ref{fig:SpeedupFPS1} and Figure~\ref{fig:SpeedupFPS2} show the results on CIFAR-10, CIFAR-100, and ImageNet dataset, respectively. 
All the results are normalized to the non-pruned ISAAC architecture when the weights are 32-bit. 
These results demonstrate that existing architectures like ISAAC and PUMA can enjoy the proposed pruning/quantization techniques since structured pruning and quantization are architecture agnostic. 
Depending upon the neural network design and dataset, applying pruning will speed up the frame processing rate of ISAAC by 7.5$\times$ to 200.8$\times$. For the PUMA, these results vary  from 5.3$\times$ up to 142$\times$.
These results also illustrate how much each proposed technique in FORMS can contribute to the frame processing rate speeding up. 
The model optimizations, including pruning, quantization, and polarization, will increase FORMS speed up by 4$\times$ to 109.6$\times$ when the fragment size is 8. When the fragment size is 16, the speed up benefits are between 5.8$\times$ up to 155.8$\times$.
As we discussed in~\ref{sec:zero-skip}, existing mixed-signal designs cannot take advantage of zero-skipping logic. FORMS utilizes this novel optimization to reach higher fps.   
By applying zero-skipping on top of model optimizations, with a fragment size of 8, the speed-up of FORMS is between 10.7$\times$ to 377.9$\times$. These results are between 11.2$\times$ up to 336.9$\times$ when FORMS uses fragment size of 16.
As demonstrated, FORMS can process frames (images) much faster than all current mixed-signal DNN accelerators.

%  \vspace{2ex}
% \vspace{1ex}

\subsection{Variation Analysis}
\label{sec:variation}
Variation is a challenge in building in-situ accelerators~\cite{dalgaty2021situ}.
We evaluate the impact of device variation of ReRAM device caused by the imperfection in fabrication technology. 
We model the device variation as a log-normal distribution~\cite{variation1}.
Table \ref{tab: variation} shows the accuracy degradation of ResNet18 (other networks have similar results) on different datasets, when introducing a device variation with 0 mean and 0.1 standard deviation. The accuracy degradation is based on the accuracy results shown in Table~\ref{table:MNIST_CIFAR} and \ref{table:CIFAR100_IMAGENET}, and it is the average value of 50 runs.
The classification task on ImageNet is much harder than CIFAR10/100. It is more sensitive to the hardware variation~\cite{mittal2020survey}. 
For instance, the original model (without adopting any compression) has around a 2.8\% accuracy degradation.
As shown in Table \ref{tab: variation}, our polarization, quantization, and zero-skipping techniques do not decrease the network robustness, while the pruning introduces a certain degradation of robustness (an extra 1.3\% accuracy degradation).
The reason is that when 50\% of the weights are pruned (under 2$\times$ pruning rate), each remaining weight becomes more important and the network resistance to variation reduces.
But it happens in all pruning works and is not a problem unique to FORMS.
The accuracy degradation can be relieved by reducing the prune ratio. 
Such a trade-off depends on the demands of real applications.
It is worth noting that the prior techniques used to improve robustness \cite{yuan2021failure,robust1,  long2019design} can be applied to FORMS.

\begin{table}[t]
\caption{Accuracy degradation caused by device variation for ResNet18 on different dataset under lognormal distribution with 0 mean and 0.1 standard deviation.}
% \vspace{-5pt}
\begin{center}
\begin{threeparttable}
\scalebox{1.0}{
\begin{tabular}{c|c|c|c|c}
\toprule
 \tabincell{c}{Dataset} & \tabincell{c}{Original \\ Model}&\tabincell{c}{Polarization\\Only}  &\tabincell{c}{Pruning\\Only}  & \tabincell{c} {Full\\Optimization}  \\
\midrule
Cifar10     &   0.35\%  & 0.37\% &   1.82\%    &   1.80\%   \\
Cifar100    &   0.72\%  & 0.68\% &   1.86\%    &   1.89\%   \\
ImageNet    &	2.87\%	& 2.86\% &   4.24\%	   &   4.21\%   \\
\bottomrule
\end{tabular}}
\end{threeparttable}
\end{center}
\label{tab: variation}
	   % \vspace{-5pt}
\end{table}

% \vspace{-5pt}
\section{Conclusion}
We propose FORMS, a fine-grained ReRAM-based DNN accelerator with algorithm/hardware co-design optimizations.
A novel fragment polarization method is proposed to elegantly solve the natural challenge of representing positive/negative weights on ReRAM crossbars without doubling crossbar cost or 
introducing extra hardware cost for result compensation.
Weight pruning and quantization technique are combined to explore a good balance between the overall hardware cost and performance.
We design a fine-grained architecture that is more robust to non-idealities and noise and makes the ADC implementation less challenging compared to the coarse-grained architecture designs.
Crucially, our novel zero-skipping logic significantly avoids unnecessary computations and reduce computation time.
Putting all together, our FORMS optimization framework can speed up ISAAC up to 377.9$\times$. 
When combining our optimization framework and architecture design, FORMS achieves 1.50$\times$ and 1.93$\times$ area and power efficiency improvements in terms of $\frac {GOPs}{s \times mm^{2}}$ and $\frac {GOPs}{W}$ over an optimized ISAAC with almost the same power/area costs.

\section*{Acknowledgement}
This work is funded by the National Science Foundation Awards CCF-1750656, CCF-1919289, CCF-1937500, and CNS-1909172.

%%%%%%% -- PAPER CONTENT ENDS -- %%%%%%%%

%%%%%%%%% -- BIB STYLE AND FILE -- %%%%%%%%

% \bibliographystyle{IEEEtran}
% \bibliography{main}
%%%%%%%%%%%%%%%%%%%%%%%%%%%%%%%%%%%%

\end{document}